\documentclass[aps,twocolumn,nofootinbib,superscriptaddress]{revtex4-1}

\usepackage[utf8]{inputenc}
\usepackage{epsfig}
\usepackage[breaklinks]{hyperref}
\usepackage{nicefrac}
\usepackage{bm}
\usepackage{dcolumn}
\usepackage{amsmath}
\usepackage{amssymb}
\usepackage{isotope}
\usepackage{xspace}
\usepackage{suffix}
\usepackage{slashed}
\usepackage{multirow}

\usepackage{color}

\begin{document}

\title{A comparison of two possible nuclear effective field theory expansions \\
  around the one- and two-pion exchange potentials}

\author{Manuel Pavon Valderrama}\email{mpavon@buaa.edu.cn}
\affiliation{School of Physics, Beihang University, Beijing 100191, China} 

\date{\today}


\begin{abstract} 
  In the effective field theory formalism nuclear forces are organized
  as a low energy expansion.
  Usually the lowest order in this expansion corresponds to the non-perturbative
  iteration of the one-pion exchange potential and a few contact-range
  operators as to ensure renormalization group invariance.
  Here we consider an alternative expansion in which two-pion
  exchange becomes the lowest order finite range interaction
  and as such is expected to be iterated.
  A comparison of this new expansion with the standard one shows a better
  convergence pattern for a few selected partial waves ($^1S_0$, $^3P_0$ and
  $^3S_1$-$^3D_1$) in two-nucleon scattering, in particular
  for the $^1S_0$ channel (though both expansions converge well).
  We briefly comment on the possible theoretical grounds for the expansion
  around two-pion exchange from the point of view of
  effective field theory.
\end{abstract}

\maketitle

\section{Introduction}

The theoretical derivation of the nuclear forces is still one of the most
pressing and interesting open problems of nuclear
physics~\cite{Machleidt:2017vls}.
Nowadays we expect any serious attempt of such a derivation to be grounded
on quantum chromodynamics (QCD), the fundamental theory of strong interactions.
Two strategies exists: a direct derivation in terms of lattice QCD~\cite{Beane:2010em},
in which QCD is solved by resorting to brute force calculations,
and an indirect one in which an effective field theory (EFT)~\cite{Bedaque:2002mn,Epelbaum:2008ga,Machleidt:2011zz}
is formulated incorporating
the low energy symmetries (most notably chiral symmetry) and
degrees of freedom (pions, nucleons and possibly deltas) of QCD.
This second strategy is in principle equivalent to QCD by virtue of
being its renormalization group evolution at low energies.

Within the EFT formulation,
we organize nuclear forces as a low energy expansion in terms of
a ratio of scales, $Q / M$, where $Q$ is a soft scale
that can be usually identified with the pion mass ($m_{\pi} \sim 140\,{\rm MeV}$)
or the typical momenta of the nucleons within nuclei and $M$ a hard scale
that might correspond to the nucleon mass $M_N$ or the chiral symmetry breaking
scale $\Lambda_{\chi} = 4 \pi f_{\pi} \sim 1\,{\rm GeV}$.
However, there is no agreement yet on how to organize the EFT expansion
for nuclear forces, as shown by the existence of numerous power counting
proposals~\cite{Weinberg:1990rz,Weinberg:1991um,Kaplan:1998tg,Kaplan:1998we,Chen:1999tn,Beane:2001bc,Nogga:2005hy,Beane:2008bt},
the eventual discovery of theoretical inconsistencies
in a few countings~\cite{Kaplan:1996xu,Fleming:1999ee,PavonValderrama:2019lsu}
and daring and original~\cite{Epelbaum:2009sd,Epelbaum:2018zli}
(but sometimes contested~\cite{PavonValderrama:2019uzi})
reinterpretations of nuclear EFT to deal with these problems.

This predicament might be explained by a poor expansion parameter ($Q/M$):
if the expansion parameter is not small enough, the relative importance of
the terms in the expansion could be very different than what
naive estimations suggest.
This might happen even if the coefficients in the expansion are of order one.
To give a concrete example, we might consider the expansion of
the matrix elements of a physical observable $\hat{\mathcal{O}}$:
\begin{eqnarray}
  \langle \hat{\mathcal{O}} \rangle =
    \sum_{\nu = \nu_{\rm min}}^{\infty} \langle \hat{\mathcal{O}}^{(\nu)} \rangle =
  \sum_{\nu = \nu_{\rm min}}^{\infty} c_{\nu} x^{\nu} \, ,
  \label{eq:expansion}
\end{eqnarray}
where $c_{\nu}$ are coefficients, $x$ the expansion parameter and
$\nu \geq \nu_{\min}$ the order in the expansion.
Provided $|x| < 1$ and that the $c_{\nu}$ do not grow exponentially
with the order $\nu$, the previous expansion
will eventually converge.

\begin{figure*}[!htt]
\begin{center}
  \includegraphics[trim={4.0cm 1.0cm 3.0cm 0.0cm},clip,height=4.75cm]{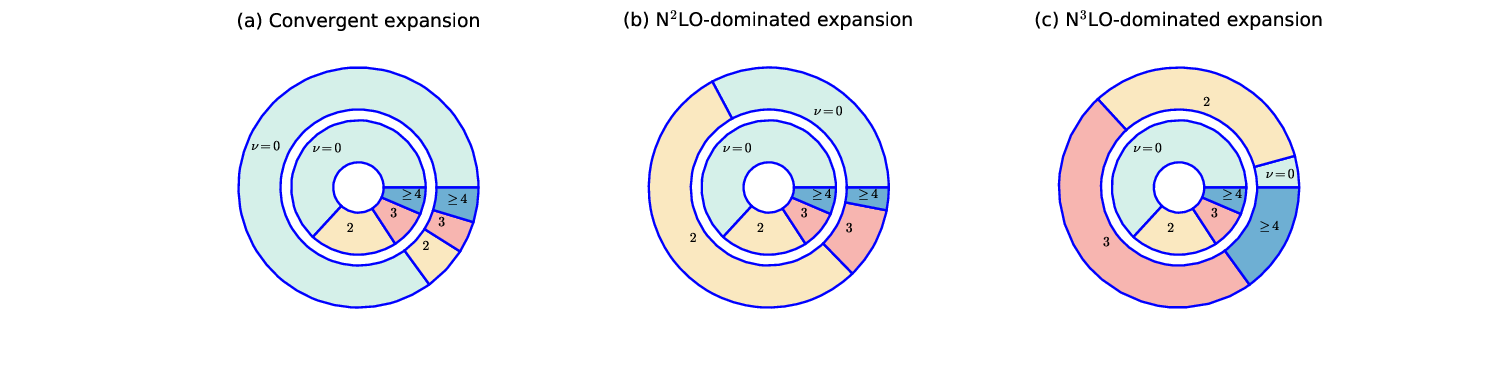}
\end{center}
\caption{
  Three possible EFT series in the toy model expansion we have proposed
  in Eqs.~(\ref{eq:expansion}) and (\ref{eq:cnu-toy}).
  The inner circle is the expected average contribution of each order
  in the EFT series, which is $\{ 1/2, 1/6, 2/27, 11/216 \}$ for
  $\nu = \{ 0, 2, 3, \geq 4 \}$ and an expansion parameter
  $x = 1/3$.
  The outer circle represents a random EFT series generated
  with our toy model, where each of the three cases shows
  a possible outcome: (a) a series converging faster than expected,
  (b) a series in which the $\nu = 2$ term happens to be larger
  than expected and (c) a series in which it is
  the $\nu = 3$ term that dominates.
  All these series are convergent, yet if we were limited to the first few
  terms we will be led to the (premature) conclusion that they are not,
  or in terms of power counting, that the power counting is failing.
  In this toy model there is no underlying reason why a particular EFT series
  happens to be {\it fine-tuned} (i.e. a subleading order contribution
  behaves as a leading order one, contrary to a priori expectations),
  it is merely a matter of chance.
  In the singlet channel of the two-nucleon system this is not necessarily
  the case: OPE cancels in the chiral limit and the contributions
  from the $\Delta$ excitations to TPE are large, favoring
  non-standard orderings analogous to (b) and (c).
  }
\label{fig:toy-counting}
\end{figure*}

However, there is significant leeway for mishaps in the previous expansion,
particularly if we are limited to its first few terms.
For instance, this is what sometimes happens with the following
toy EFT expansion of the observable $\langle \mathcal{O} \rangle$,
in which the $c_{\nu}$ coefficients ($\nu_{\rm min} = 0$)
are given by a random value within the interval:
\begin{eqnarray}
  c_{\nu} \in [0,\nu+1] \quad
  \mbox{for $\nu \neq 1$ (and $c_{1} = 0$)} \, , \label{eq:cnu-toy}
\end{eqnarray}
where the coefficients grow with the order, to mimic the increase
in the number of diagrams. We have also set the $\nu = 1$
coefficient to zero, a choice which resembles nuclear EFT
as applied to the two-nucleon forces when naive
dimensional analysis (NDA) is followed.
For simplicity we have taken all coefficients to be positive.
With the previous set of coefficients and an expansion parameter of $x = 1/3$
(representative of the situation in the two-nucleon sector),
the convergence pattern of the observable $\hat{\mathcal{O}}$ is
\begin{alignat}{6}
  \overline{\langle \hat{\mathcal{O}} \rangle} =
  &\overline{\langle \hat{\mathcal{O}}^{(0)} \rangle}& +
  &\overline{\langle \hat{\mathcal{O}}^{(1)} \rangle}& +
  &\overline{\langle \hat{\mathcal{O}}^{(2)} \rangle}& +
  &\overline{\langle \hat{\mathcal{O}}^{(3)} \rangle}& &+
  \sum_{\nu \geq 4} &\overline{\langle \hat{\mathcal{O}}^{(\nu)} \rangle} \\
  =
  &\quad \frac{1}{2}& + &\quad 0& + &\quad \frac{1}{6}&  +
  &\,\,\,\, \frac{2}{27}& &+ & \frac{11}{216} \, ,
\end{alignat}
where the bar indicates the expected value (equivalent to taking
$c_{\nu} = (\nu+1)/2$ for $\nu \neq 1$) and the second line
lists the magnitude of each contribution.
Yet, while convergence is guaranteed for this toy expansion, the actual
convergence pattern depends on chance and it can significantly
deviate from the previous expectations.
If we generate random toy expansions with $x=1/3$ and the $c_{\nu}$
coefficients in Eq.~(\ref{eq:cnu-toy}) most of them will follow
the expected power counting pattern
\begin{eqnarray}
  | \overline{\langle \hat{\mathcal{O}}^{(0)} \rangle} | >
  | \overline{\langle \hat{\mathcal{O}}^{(2)} \rangle} | >
  | \overline{\langle \hat{\mathcal{O}}^{(3)} \rangle} | > \dots \, ,
\end{eqnarray}
but a few will not.
This is illustrated in Fig.~\ref{fig:toy-counting}, where three random
toy expansions generated from the previous rules are shown.
One of them is more convergent than expected,
but the other two are dominated by the $\nu=2$ ($\rm N^2LO$) and
$\nu=3$ ($\rm N^3LO$) terms, respectively.
In these two cases and if we consider the first few terms only,
it will be as if the EFT expansion is not working,
despite the fact that it is completely convergent.

  The bottom-line is that for systems where the convergence parameter
  is not small enough, the probability that the expansion
  is in disarray is sizable.
  Terms expected to be subleading might behave as leading,
  as if the EFT expansion itself has been {\it fined-tuned}.
  This by itself does not invalidate the idea of the EFT expansion
  or its eventual convergence, but it will certainly mess up with its practical
  implementation, particularly if we are not willing to go to arbitrarily
  high orders for checking whether there is convergence of not.
  Luckily, this situation can be dealt with by rearranging the power counting
  at lowest orders, thus avoiding the inconvenience of having to calculate
  high orders of the expansion.
  For the toy model we have considered here the only reason behind unexpected
  behaviors of the EFT expansion is chance, but for other physical systems
  there might be underlying reasons maybe related to neglected
  light scales.
  Also, in contrast with toy EFT series, for actual physical systems
  it is imperative to properly renormalize the theory before reaching
  any conclusions about the behavior of their expansions.

Something in the previous line is probably happening in the two-nucleon system.
If we expand the finite-range part of the effective potential derived
from chiral perturbation theory (i.e. the pion exchanges),
we have
\begin{eqnarray}
  V_{\rm NN}^{\rm EFT}(r > 0) &=& \sum_{\nu = 0}^{\infty}\,V^{(\nu)}(r) \nonumber \\
  &=& \underbrace{V^{(0)}(r)}_{V_{\rm OPE}} +
  \underbrace{V^{(2)}(r) + V^{(3)}(r)}_{V_{\rm TPE}} + \dots \, ,
\end{eqnarray}
where the superscript indicates the order of a contribution in naive dimensional
analysis (NDA) and below them we specify whether this contribution is
one-pion-exchange (OPE) or (mostly~\footnote{The $\nu \geq 2$ pieces
  also contain small corrections to the OPE potential,
  but for simplicity we will ignore them here.})
two-pion-exchange (TPE), which can be further subdivided
into leading ($\nu = 2$) and subleading ($\nu = 3$) TPE,
or TPE(L) and TPE(SL) for short.
Naively, we expect the behavior of the $Q^{\nu}$ contributions to be
\begin{eqnarray}
  V^{(\nu)}(r) \sim \frac{4 \pi}{M_N}\,
  \frac{1}{{\left[ \Lambda_{\rm NN}^{(\nu)} r \right]}^{\nu+1}}
  \,\frac{1}{r^{2}}\,v^{(\nu)}(m_{\pi} r) \, , \label{eq:VF-naive}
\end{eqnarray}
where the characteristic scale $\Lambda_{\rm NN}^{(\nu)}$ is expected to be a hard
scale, i.e. $\Lambda_{\rm NN}^{(\nu)} \sim M$ and with $v^{(\nu)}$ a function
(usually such that $\lim_{x \to 0} x\,v^{(\nu)}(x) = 0$ to avoid a modification
of the power-law behavior for $m_{\pi} r \to 0$ coming from $v^{(\nu)}$)
that captures the features of the order $\nu$ contribution to
the potential beyond its leading power-law behavior.

But, as it often happens, reality does not match expectations.
If we consider the chiral limit ($m_{\pi} \to 0$) for simplicity,
the ${}^1S_0$ and ${}^3S_1$ partial wave potentials
can be expanded as follows
\begin{eqnarray}
  \lim_{m_{\pi} \to 0 } V^{\rm EFT}_{\rm NN}(r) &=&
  - \frac{4 \pi}{M_N}\left[ \frac{1}{\Lambda_{\rm TPE(L)}^3} \frac{1}{r^5}
    + \frac{1}{\Lambda_{\rm TPE(SL)}^4} \frac{1}{r^6} \right]   \nonumber \\
  &+& \mathcal{O}\left( {(\frac{Q}{M})}^4 \right) \, ,
  \label{eq:V-chiral}
\end{eqnarray}
where in this limit the OPE contribution vanishes for S-waves and the $1/r^5$
and $1/r^6$ contributions correspond to the $\nu = 2$ and $3$ terms
in the EFT expansion of the potential, i.e. TPE(L) and TPE(SL).
From direct inspection of the potentials~\cite{Rentmeester:1999vw},
we find that for the $c_3 = -3.4\,{\rm GeV}$ and $c_3 = +3.4\,{\rm GeV}$
choice of chiral couplings the actual numerical values of
the leading and subleading TPE scales are
\begin{eqnarray}
  \Lambda_{\rm TPE(L)}(^1S_0) &=& +389\, {\rm MeV} \, , \label{eq:LambdaTPE-1} \\
  \Lambda_{\rm TPE(SL)}(^1S_0) &=& +233\, {\rm MeV} \, , \label{eq:LambdaTPE-2} \\
  \nonumber \\
  \Lambda_{\rm TPE(L)}(^3S_1) &=& -370\, {\rm MeV} \, , \\
  \Lambda_{\rm TPE(SL)}(^3S_1) &=& +220\, {\rm MeV} \, ,
  \label{eq:LambdaTPE-4}
\end{eqnarray}
which for the case of subleading TPE significantly differs from the NDA
expectation of $\Lambda_{\rm TPE} \sim (0.5-1.0)\,{\rm GeV}$.

Of course, the previous observation does not necessarily imply that
the TPE potential has to be iterated (except maybe in the chiral limit).
For physical pion masses, the TPE potential is suppressed by a $e^{-2 m_{\pi} r}$
factor at large distances, which for sure attenuates its strength.
Besides, if tensor OPE is present (as happens in the triplets),
the TPE potential will be further attenuated owing to the distortions
generated by the non-perturbative tensor OPE wave function.
Be it as it may, it will be interesting to explore what happens when TPE
is non-perturbative, particularly in comparison to what happens
when it is a perturbation.

Actually, this idea was already explored
in the past~\cite{PavonValderrama:2019lsu,Valderrama:2010aw},
though usually not in the context of building a serious EFT expansion around it
(except in~\cite{Mishra:2021luw}).
In the Weinberg prescription, scattering amplitudes are obtained from iterating
the full effective potential with a finite cutoff.
As has been often pointed out, this prescription does in general not generate
renormalizable scattering amplitudes~\cite{Kaplan:1996xu,Nogga:2005hy},
which thus might very well break the power counting assumptions of
the effective potential from which they are calculated.
Indeed, from analyzing the underlying power counting of the scattering
amplitudes calculated in the Weinberg prescription by trial and error,
it is easy to find that the Weinberg ${}^1S_0$ phase shifts are well
reproduced if TPE (plus a contact operator to guarantee
renormalizability) is ${\rm LO}$ and everything else
is subleading
(and perturbative)~\cite{PavonValderrama:2019lsu,Valderrama:2010aw}.
Recently Ref.~\cite{Mishra:2021luw} has considered this expansion
as a possible organizing principle for the $^1S_0$ singlet,
leading to results which we agree with here.

For exploring the expansion around TPE, we will do as follows: first,
we will consider the type of non-relativistic EFT expansion
that is generated from an arbitrary choice of a leading order
potential, which we will then particularize for the cases in
which OPE and TPE are leading (Sect.~\ref{sec:counting});
after this, we will calculate the EFT expansion of the phase shifts
for the two cases (Sect.~\ref{sec:two-expansions});
and finally, we will discuss and compare
these two EFTs, the possible justifications of the TPE expansions
and its implications (Sect.~\ref{sec:discussion}).
In addition, a few of the technical details of the EFT expansion
are discussed in Appendix \ref{app:expansion},
while Appendix \ref{app:scaling} analyzes the scaling
properties of this expansion, which might have applications
for the study of EFT convergence.

\section{Power counting as a function of the choice of a leading order}
\label{sec:counting}

Now we explore the possible form of the power countings generated
from expanding around the cases in which OPE and TPE are
non-perturbative.
For concreteness we will refer to them as EFT(OPE) and EFT(TPE).
Actually, EFT(OPE) has already been extensively discussed
in the literature~\cite{Birse:2005um,Valderrama:2009ei,Valderrama:2011mv,Long:2011qx,Long:2011xw,Long:2012ve}
and we will rely on preexisting analyses up to a certain extent.
The expansion of interest here is EFT(TPE), which will require modifications to
the power counting that are genuinely different from those of EFT(OPE).

For simplicity, we will consider a general EFT expansion that at ${\rm LO}$
requires the non-perturbative iteration of a piece of the effective
potential, which in turns generates a ${\rm LO}$ wave function,
$\Psi_{\rm LO}$, the form of which we assume to be known.
The reason for adopting this approach is that 
the power counting of the contact-range potential is indeed determined
by the power-law properties of $\Psi_{\rm LO}$~\cite{Valderrama:2014vra}.
Then, by particularizing $\Psi_{\rm LO}$ to the EFT(OPE) and EFT(TPE) expansions
we will deduce their power countings.

Before beginning, there are a few conventions we follow: the counting of
the non-relativistic potential will be defined with respect to its scaling
in momentum space, e.g. a potential that has the form
\begin{eqnarray}
  V(\vec{q}) \propto
  \frac{1}{M^{n+2} Q^m}\,P_{n}(\vec{q})\,v(\frac{\vec{q}}{Q}) \, ,
  \label{eq:V-counting}
\end{eqnarray}
where $\vec{q}$ is the exchanged momentum, $P_{n}$ a polynomial of order $n$
and $v(\vec{x})$ some other (non-polynomial) function,
will be counted as $Q^{n-m}$.
When considering the iterations of the potential, i.e. $V G_0 V \dots G_0 V$,
the propagator $G_0$ between two instances of the potential is counted
as $Q$.
As a consequence, if a part of the potential is treated non-perturbatively,
this is because we consider it to scale as $Q^{-1}$ or ${\rm LO}$.
Also, the notation $Q^{\nu}$ is actually short-hand for $(Q/M)^{\nu}$.

We will assume the following power-law behavior for the ${\rm LO}$ wave function
\begin{eqnarray}
  \langle r | \Psi_{\rm LO} \rangle \sim r^{\alpha-1} \, ,
\end{eqnarray}
where the exponent $\alpha$ can be calculated
from the form of the ${\rm LO}$ potential.
Following~\cite{Valderrama:2014vra} we will consider
the matrix elements of a contact-range interaction of the type
\begin{eqnarray}
  V_C(r) = \frac{C(R_c)}{4 \pi R_c^2}\,\delta(r-R_c) \, ,
\end{eqnarray}
which for concreteness we have regularized with a delta-shell,
where $R_c$ is a cutoff radius.
This contact will be sandwiched between the LO wave functions, leading to
\begin{eqnarray}
  \langle \Psi_{\rm LO} | V_C | \Psi_{\rm LO} \rangle \sim
  \frac{C(R_c)}{R_c^{2-2 \alpha}} \, . \label{eq:VC-matrix}
\end{eqnarray}
If we demand renormalization group (RG) invariance for this matrix element
\begin{eqnarray}
  \frac{d}{d R_c} \left[ \frac{C(R_c)}{R_c^{2-2\alpha}} \right] = 0 \, ,
  \label{eq:RG-contact}
\end{eqnarray}
then we realize that the running of $C(R_c)$ is given by
\begin{eqnarray}
  C(R_c) \propto R_c^{2 - 2 \alpha} \, .
\end{eqnarray}
If we evolve $C(R_c)$ from $M R_c \sim 1$ (where NDA applies and
$C(R_c) \sim 1/M^2$~\cite{Valderrama:2016koj}) to $Q R_c \sim 1$,
we find that at low resolutions this coupling behaves as
\begin{eqnarray}
  C(R_c \sim 1/Q) \sim \frac{1}{M^2}\,{\left(\frac{M}{Q}\right)}^{2-2\alpha} \, .
\end{eqnarray}
That is, if $\alpha > 1$ ($\alpha < 1$) then $C(R_c)$ will be demoted
(promoted) to order $Q^{2\alpha - 2}$.
If we further expand the coupling in energy/momentum
\begin{eqnarray}
  C = C_0 + C_2\,k^2 + C_4\,k^4 + \dots \, ,
\end{eqnarray}
it is apparent that $C_2$, $C_4$, $\dots$ are further suppressed by
factor of $Q^2$, $Q^4$, $\dots$ and scale as
\begin{eqnarray}
  C_2 \propto Q^{2 \alpha} \quad , \quad C_4 \propto Q^{2 \alpha+2} \, \quad
  \mbox{and so on.}
\end{eqnarray}

As a side note the previous argument can be cross-checked in two different ways:
finiteness of the subleading order calculations~\cite{Valderrama:2009ei,Valderrama:2011mv} and residual cutoff dependence~\cite{Griesshammer:2015osb,Griesshammer:2020fwr,Odell:2021lia}.
If we consider the perturbative correction of a hypothetical, unaccounted for
$Q^{\nu}$ contribution to the chiral potential in EFT, the matrix
element will behave as~\cite{Valderrama:2016koj}
\begin{eqnarray}
  \langle \Psi_{\rm LO} | V^{(\nu)} | \Psi_{\rm LO} \rangle
  \sim \int_{R_c}^{\infty} \frac{dr}{r^{3+\nu}} r^{2 \alpha}\,
  (1 + k^2 r^2 + \dots )  \, ,
  \nonumber \\ \label{eq:VF-matrix}
\end{eqnarray}
where $V^{(\nu)}$ refers to a finite-range contribution of order $\nu$ to
the EFT potential.
From this, the lowest order at which we encounter a divergence in EFT
is $Q^{2\alpha - 2}$ (which can be absorbed by a recalibration of
the scattering length), while at $Q^{2 \alpha}$ we will find a
divergence that requires a range correction.
From the point of view of residual cutoff dependence, the LO phase shift
converges as~\cite{PavonValderrama:2007nu}
\begin{eqnarray}
  \frac{d}{d R_c} \delta_{\rm LO}(k; R_c) \propto k^3 R_c^{2\alpha + 1} \, ,
\end{eqnarray}
for $R_c \to 0$, which suggest that the range corrections enter $Q^{2\alpha+1}$
orders after ${\rm LO}$ ($Q^{-1}$).

At this point it has to be stressed that two of the previous arguments
(Eqs.~(\ref{eq:VC-matrix}) and (\ref{eq:VF-matrix}))
are perturbative in nature and do not take into account that
the LO wave function might already require a series of
contact interaction for its unambiguous determination.
Indeed, this is what happens when the ${\rm LO}$ wave function
is an attractive singular
interaction~\cite{PavonValderrama:2005wv,PavonValderrama:2005uj}.
If this is the case, the necessary contact-range interactions
will be automatically promoted to ${\rm LO}$ or $Q^{-1}$.
If we were to use the language of
Refs.~\cite{Birse:1998dk,Barford:2002je,Birse:2005um},
we will say that the ${\rm LO}$ contact-range interactions
are part of the RG fixed point, while the counting derived from
Eqs.~(\ref{eq:VC-matrix}) and (\ref{eq:VF-matrix}) corresponds
to that of the perturbations around said fixed point.

Now, all that remains is to determine the exponent $\alpha$, which is a
relatively well-known quantity:
\begin{itemize}
\item[(i)] For a ${\rm LO}$ regular potential, we have $\alpha=0$ in S-waves.
\item[(ii)] For a ${\rm LO}$ power-law singular potential
  of the type $1/r^{n}$ with $n \geq 2$, we have $\alpha = n/4$. 
\end{itemize}
With this, the power counting of EFT(OPE) can be reproduced by taking into
account that in the $^1S_0$ singlet $\alpha = 0$, while for the triplets
in which tensor OPE is non-perturbative (e.g. $^3S_1$-$^3D_1$ and $^3P_0$),
$\alpha = 3/4$.

In contrast for EFT(TPE) we have $\alpha = 3/2$
for singlets and triplets alike.
This implies in particular that the $C_2$ couplings, i.e. range corrections,
are all demoted to $Q^3$ or ${\rm N^4LO}$.
This demotion of the range corrections is probably the most characteristic
feature of the power countings arising from expanding around
singular interactions.

In Table \ref{tab:counting} we briefly summarize the power counting of
EFT(OPE) and EFT(TPE) as will be implemented and explored in this work.
It includes a few simplifying assumptions, e.g. the fractional counting
of $C_2$ with tensor OPE (i.e. $Q^{3/2}$ from $\alpha = 3/4$)
has been approximated to $Q^2$ and a few couplings that
are numerically small (the couplings that fix the $E_1$
and $^3D_1$ scattering lengths)
have been demoted from $Q^{-1/2}$ to $Q^2$.

The most important simplification in EFT(TPE) is that we have simply
promoted the full $\nu \leq 3$ EFT potential to ${\rm LO}$ (instead of
only promoting subleading TPE, which makes more sense in view of
$\Lambda_{\rm TPE(SL)}$, check Eqs.~(\ref{eq:LambdaTPE-1}-\ref{eq:LambdaTPE-4})).
This might be justified from the observation that the contributions of OPE
and leading TPE are numerically small.
Yet, the actual reason why it is safe to promote them is that this choice
does not change the RG evolution (RGE) of EFT(TPE): the power-law behavior of
OPE and TPE(L) is unable to change the RGE of a theory
in which TPE(SL) is treated non-perturbative.
Besides, perturbative OPE is finite when sandwiched between the LO wave
functions of EFT(TPE), while TPE(L) is renormalized by simply
recalibrating the scattering length, i.e. by a subleading
correction to $C_0$.

\begin{table}
\begin{center}
\begin{tabular}{|c|c|c|c|}
  \hline \hline
  Contribution & NDA & EFT(OPE) & EFT(TPE) \\
  \hline
  $V_{\rm OPE}$ & $Q^0$ & $Q^{-1}$ & $Q^{-1}$ \\
  $V_{\rm TPE(L)}$ & $Q^2$ & $Q^2$ & $Q^{-1}$\\
  $V_{\rm TPE(SL)}$ & $Q^3$ & $Q^3$ & $Q^{-1}$\\
  \hline
  $C_0\,({}^1S_0)$ & $Q^0$ & $Q^{-1}$ & $Q^{-1}$ \\
  $C_2\,({}^1S_0)$ & $Q^2$ & $Q^{0}$ & $Q^3$ \\
  $C_4\,({}^1S_0)$ & $Q^4$ & $Q^{2}$ & $Q^5$ \\
  \hline
  $C_0\,({}^3S_1)$ & $Q^2$ & $Q^{-1}$ & $Q^{-1}$\\
  $C_0\,(E_1 / {}^3D_1)$ & $Q^2$ & $Q^{2}$ & $Q^{-1}$\\
  $C_2\,({}^3S_1 / E_1 / {}^3D_1)$ & $Q^4$ & $Q^{2}$ & $Q^3$\\
  \hline
  $C_0\,({}^3P_0)$ & $Q^0$ & $Q^{-1}$ & $Q^{-1}$ \\
  $C_2\,({}^3P_0)$ & $Q^2$ & $Q^{2}$ & $Q^3$ \\
  \hline \hline
\end{tabular}
\end{center}
\caption{
  Power counting for the different contributions to the effective potential
  in the EFT(OPE) and EFT(TPE) expansions as defined in this work and
  their comparison with NDA estimations.
  The EFT(OPE) counting has been extensively studied in
  Refs.~\cite{Birse:2005um,Valderrama:2016koj,Valderrama:2009ei,Valderrama:2011mv,Long:2011qx,Long:2011xw,Long:2012ve},
  where there exists differences in the details of how the different
  operators are counted (esteeming from the diverging assumptions
  made in each of the previous references).
  Here, while EFT(TPE) counting is purely driven by RG invariance,
  a few of the choices we have made in EFT(OPE) are practical in nature:
  in the ${}^3S_1$-${}^3D_1$ channel the couplings fixing the scattering
  volumes and hypervolumes in the $E_1$ and ${}^3D_1$ partial waves
  ($C_0(E_1)$ and $C_0({}^3D_1$) in the Table)
  are demoted to order $Q^2$, even though they should enter at
  order $Q^{-1/2}$ if we follow RG analysis~\cite{Birse:2005um}.
  For the singlet we follow Ref.~\cite{Valderrama:2016koj}.
}
\label{tab:counting}
\end{table}

\section{Comparison between the two expansions}
\label{sec:two-expansions}

Having decided the two power countings, the only thing left is to make
the necessary calculations in the EFT(OPE) and EFT(TPE) expansions.
The technical details are straightforward but tedious,
and can be consulted in Appendix \ref{app:expansion},
where here we simply indicate the main features of the calculation:
\begin{itemize}
\item[(i)] The LO wave functions and scattering amplitudes are obtained and
  renormalized as in Refs.~\cite{PavonValderrama:2005gu,PavonValderrama:2005wv,PavonValderrama:2005uj}, in which by using suitable boundary conditions
  it was shown the minimum number of couplings required
  for the different partial waves.
  In particular for the coupled channels (e.g. ${}^3S_1-{}^3D_1$) we have that:
  \begin{itemize}
  \item[(i.a)] Non-perturbative OPE can be renormalized by fixing one of
    the scattering lengths (or their L-wave
    equivalents)~\cite{PavonValderrama:2005gu}.
  \item[(i.b)] Non-perturbative TPE can be renormalized by fixing the three
    scattering lengths (provided that the two eigenvalues of the
    coupled channel potentials are attractive, which is what happens
    in most partial waves)~\cite{PavonValderrama:2005wv,PavonValderrama:2005uj}.
  \end{itemize}

\item[(ii)] For the subleading order phase shifts, we basically follow the
  regularization used in Refs.~\cite{Valderrama:2009ei,Valderrama:2011mv},
  which also shows that the perturbative corrections are indeed
  renormalizable after the inclusion of a series of couplings.
  A few differences with respect to Refs.~\cite{Valderrama:2009ei,Valderrama:2011mv} are worth commenting:
  \begin{itemize}
  \item[(ii.a)] For the ${}^1S_0$ partial wave in EFT(OPE),
    Refs.~\cite{Valderrama:2009ei,Valderrama:2011mv}
    included both the $C_2$ and $C_4$ couplings at $Q^2$,
    which is the minimal requirement to render perturbative TPE finite and
    well-defined in the $R_c \to 0$ limit.
    However, finiteness is merely a subset of renormalizability and indeed
    RG invariance requires the $C_2$ coupling to enter at order $Q^0$,
    which is what we do here, a choice that entails a series of iterations of
    the $C_2$ coupling (which is computationally fastidious and has prompted
    expansions in which $C_2$ is fully iterated
    at ${\rm LO}$~\cite{Peng:2021pvo}).
    This is further justified by the fact that the ${\rm LO}$ calculation
    of the $^1S_0$ effective range only reproduces $50\%$ of its value.

  \item[(ii.b)] For the ${}^3S_1$-${}^3D_1$ channel in EFT(OPE),
    Refs.~\cite{Valderrama:2009ei,Valderrama:2011mv} use a total of
    six couplings for renormalizing TPE. However, Long and Yang
    proved that three couplings are enough~\cite{Long:2011xw}
    (just as happens in Weinberg's counting). We nonetheless will
    use six here, as this provides a more direct comparison
    between EFT(OPE) and EFT(TPE) at $\rm N^4LO$
    (as with this choice we will end up with six couplings
    in both expansions).

  \item[(ii.c)] This choice of six couplings generates a further complication
    for ${}^3S_1$-${}^3D_1$ in EFT(OPE): RGE arguments (either
    Eq.~(\ref{eq:RG-contact}) particularized for $\alpha = 3/4$
    or Ref.~\cite{Birse:2005um} for its original formulation)
    indicate that the couplings fixing the scattering lengths
    in the $E_1$ and $^3D_1$ channels enter at order $Q^{-1/2}$
    or ${\rm N^{1/2}LO}$. This is problematic as it will require
    an inordinate number of iterations of these couplings (far
    exceeding the required iterations of $C_2$ in the singlet).
    Yet, it is worth noticing that non-perturbative OPE actually
    reproduces relatively well these scattering lengths
    at ${\rm LO}$~\cite{PavonValderrama:2005gu} (in contrast
    with the $^1S_0$ case, in which this does not happen
    for the effective range),
    indicating that in most practical settings these couplings are
    not really required. For this reason we will simply demote
    the $C_0(E_1 / {}^3D_1)$ couplings to $Q^2$.

  \item[(ii.d)] In principle a similar problem appears in the $^3P_0$ partial
    wave, but in this case the scattering length can be reproduced at
    $\rm LO$, which automatically implies that the subleading
    corrections to $C_0({}^3P_0)$ are trivial.

  \item[(ii.e)] Finally, in EFT(OPE) we further assume that the promotion of
    OPE from its naive scaling ($Q^0$) to ${\rm LO}$ does not affect
    the power counting of the TPE potential. This is in contrast
    to the approach of Long and Yang which also promotes
    the TPE potential by one order~\cite{Long:2011qx,Long:2011xw,Long:2012ve},
    a choice with makes perfect sense in terms of scales once we look
    at Eqs.~(\ref{eq:LambdaTPE-1}-\ref{eq:LambdaTPE-4})).
  \end{itemize}
\end{itemize}

Having explained the previous details, we might simply proceed to
the calculation of the $^1S_0$, $^3P_0$ and $^3S_1$-$^3D_1$
phase shifts in EFT(OPE) and EFT(TPE).
For this, regardless of the expansion, we will calibrate the ${\rm LO}$
couplings as to reproduce the following values of the scattering
length: $\alpha_0({}^1S_0) = -23.7\,{\rm fm}$,
$\alpha_0({}^3P_0) = -2.65\,{\rm fm}^3$, $\alpha_0({}^3S_1) = 5.42\,{\rm fm}$,
$\alpha_0(E_1) = 1.67,{\rm fm}^3$ and $\alpha_0({}^3D_1) = 6.60\,{\rm fm}^5$,
where the $E_1$ and ${}^3D_1$ scattering lengths are only required
for EFT(TPE) and are defined for the Stapp-Ypsilantis-Metropolis
(SYM, also known as nuclear bar) parametrization of
the phase shifts~\cite{PhysRev.105.302} (where their detailed low energy
behavior can be consulted in Ref.~\cite{PavonValderrama:2004se}).
For the subleading order phase shifts, the couplings will be determined from
fitting the Nijmegen II phase shifts~\cite{Stoks:1994wp} (expected
to be equivalent to the ones extracted in the Nijmegen PWA~\cite{Stoks:1993tb}
within errors) within the $k = (100-200)\,{\rm MeV}$
center-of-mass momentum window, except for the singlet channel where
the momentum window changes with the order and includes lower
momenta to ensure the correct scattering length (in particular
we use $k = (10-40)\,{\rm MeV}$, $(10-80)\,{\rm MeV}$,
$(10-200)\,{\rm MeV}$ and $(10-200)\,{\rm MeV}$ at 
${\rm NLO}$, ${\rm N^2LO}$, ${\rm N^3LO}$ and ${\rm N^4LO}$).
The explicit expressions for the effective OPE and TPE potential are taken
from Ref.~\cite{Rentmeester:1999vw}, where for the couplings
we follow the same choices as in Ref.~\cite{Valderrama:2011mv}
--- that is, $g_A = 1.26$, $f_{\pi} = 92.4\,{\rm MeV}$,
$m_{\pi} = 138.03\,{\rm MeV}$, $d_{18} = -0.97\,{\rm GeV}^2$ (equivalent
to using $g_A=1.29$ instead of $1.26$ in the OPE piece of the potential),
$c_1 = -0.81\,{\rm GeV}^1$, $c_3 = -3.4\,{\rm GeV}^{-1}$ and
$c_4 = +3.4\,{\rm GeV}^{-1}$ --- and where the recoil (or
$1/M_N$) corrections are included.

\begin{figure*}[ttt]
\begin{center}
  \includegraphics[width=4.90cm]{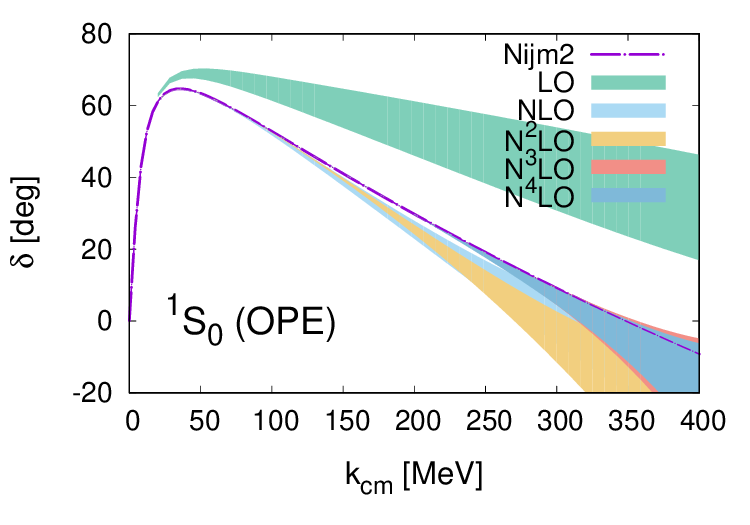}
  \includegraphics[width=4.90cm]{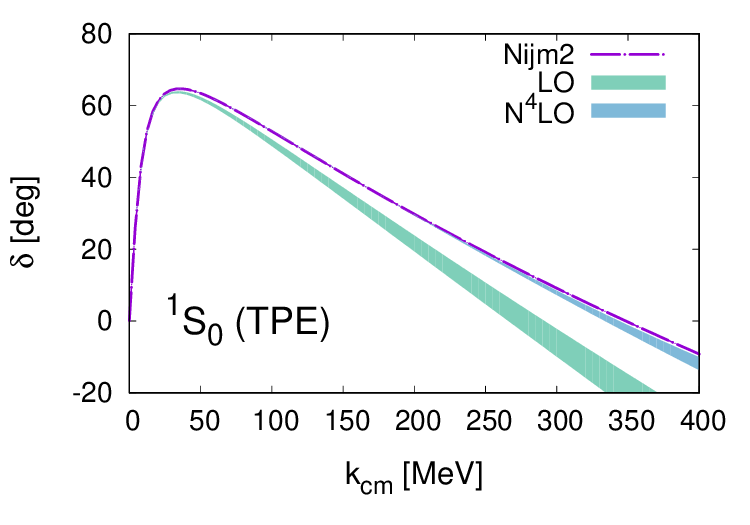}
  \includegraphics[width=4.90cm]{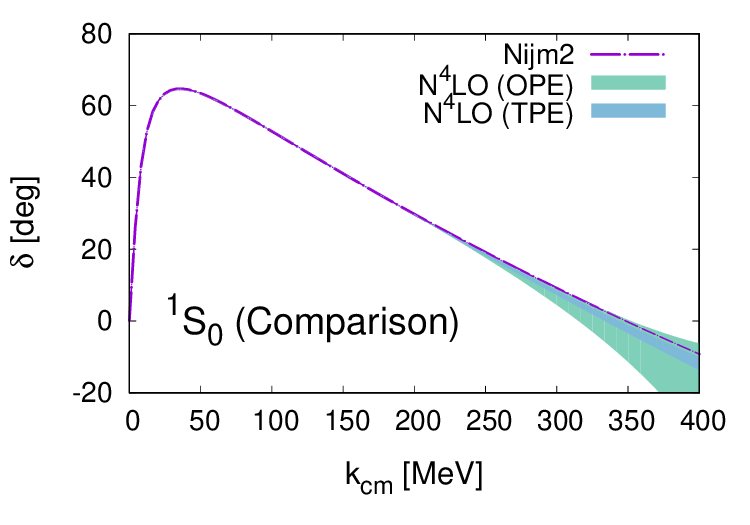}
  \includegraphics[width=4.90cm]{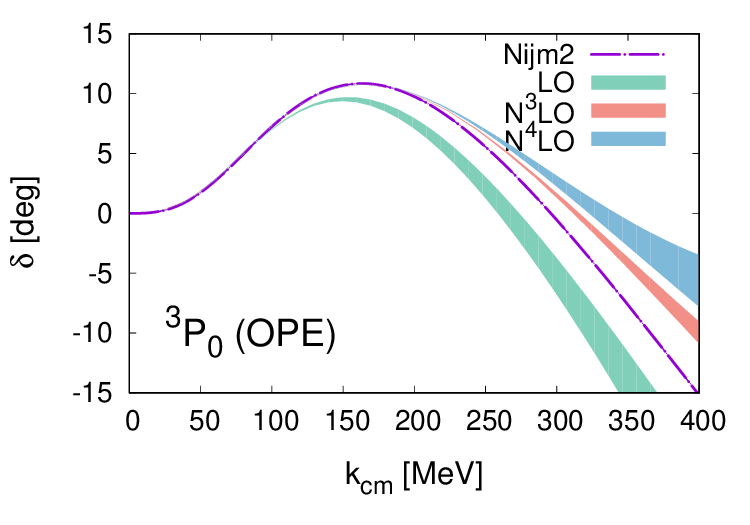}
  \includegraphics[width=4.90cm]{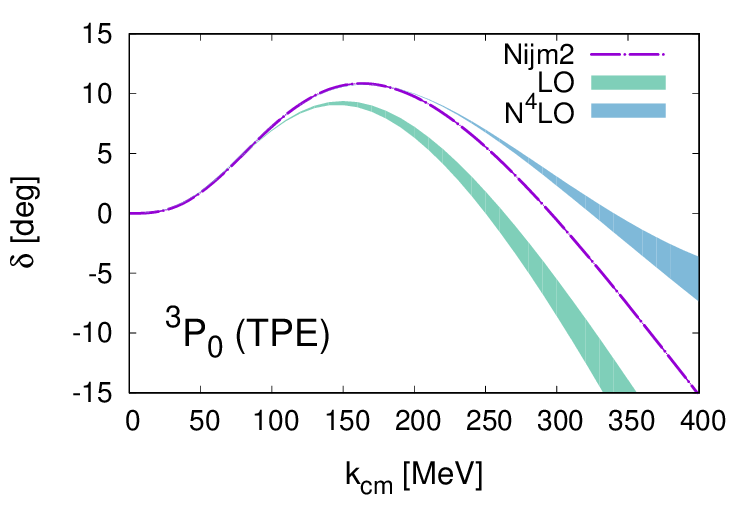}
  \includegraphics[width=4.90cm]{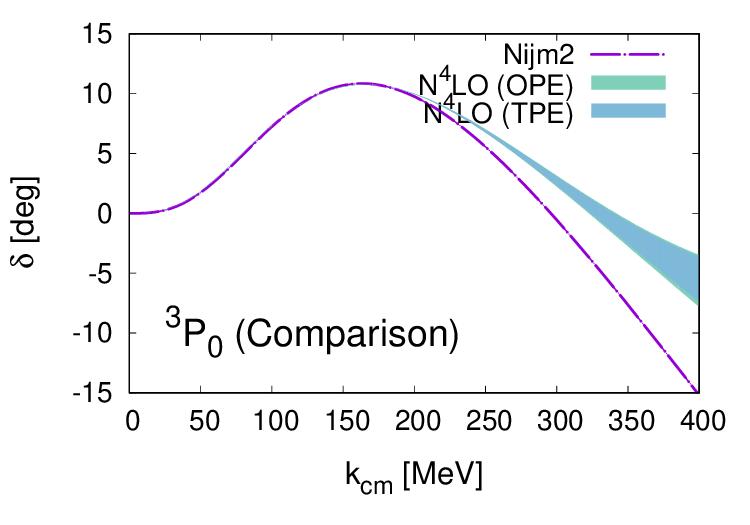}
  \includegraphics[width=4.90cm]{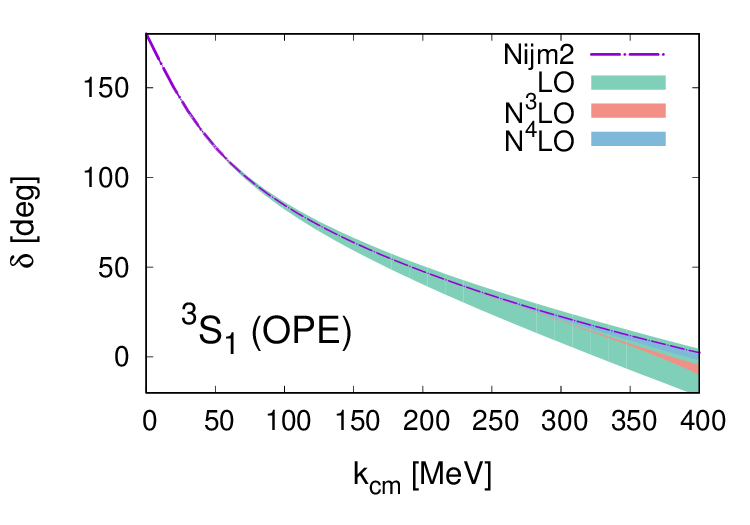}
  \includegraphics[width=4.90cm]{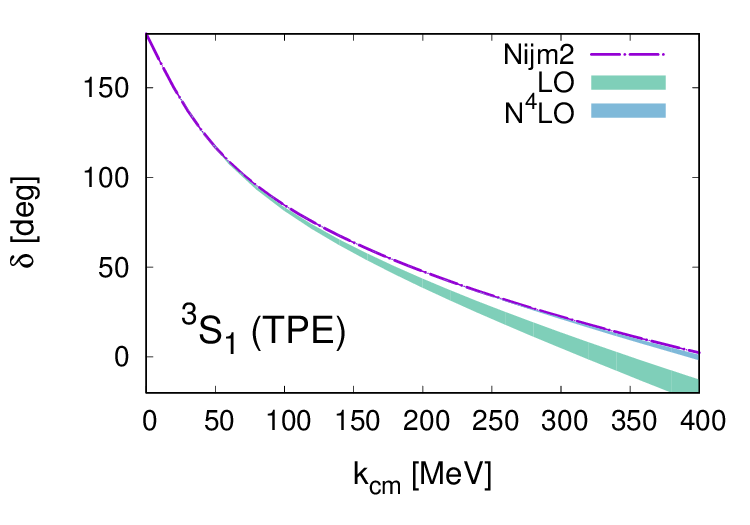}
  \includegraphics[width=4.90cm]{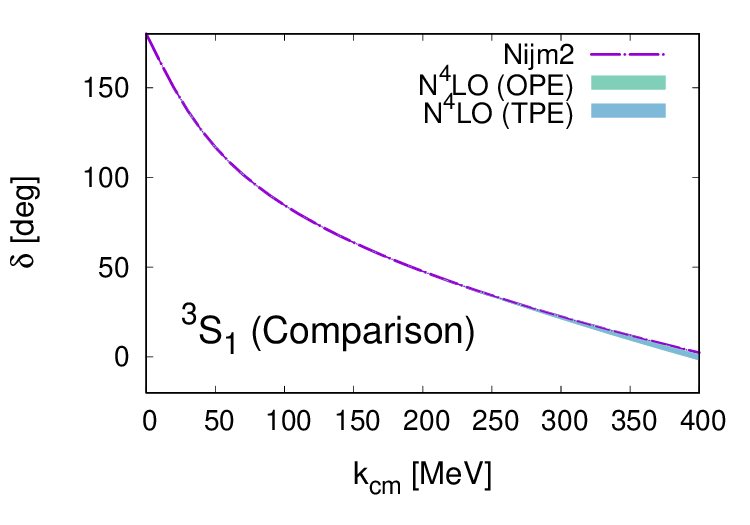}
  \includegraphics[width=4.90cm]{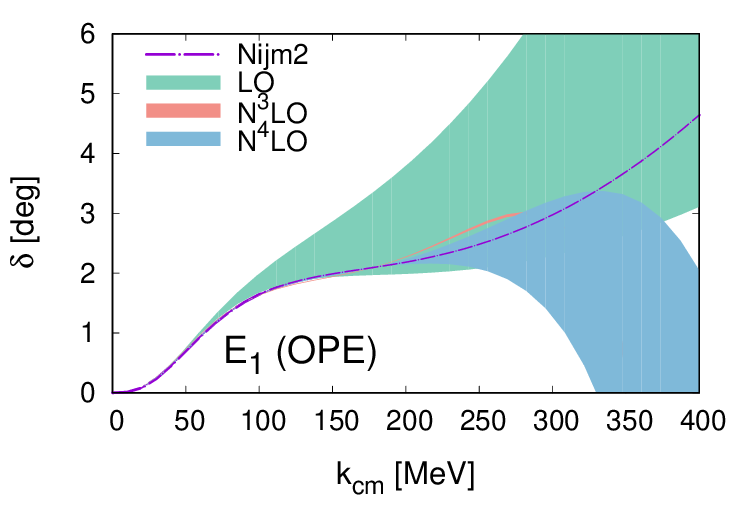}
  \includegraphics[width=4.90cm]{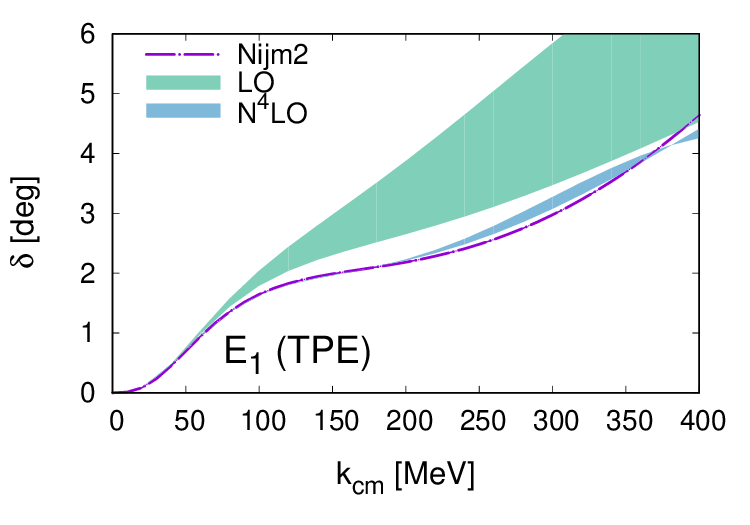}
  \includegraphics[width=4.90cm]{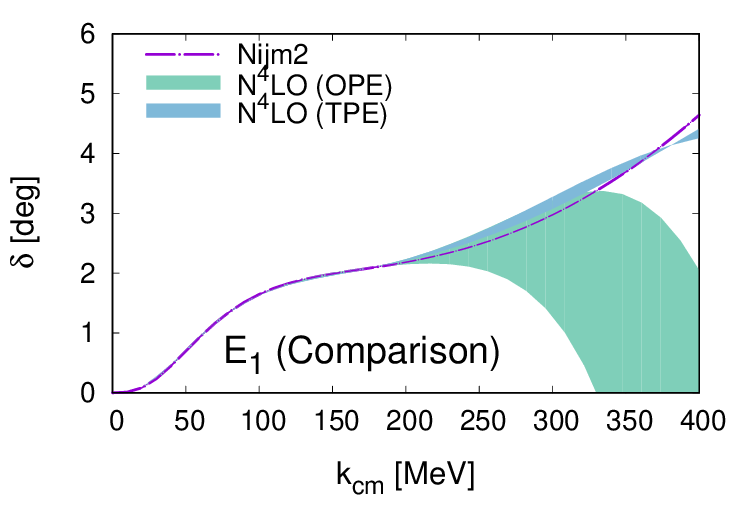}
  \includegraphics[width=4.90cm]{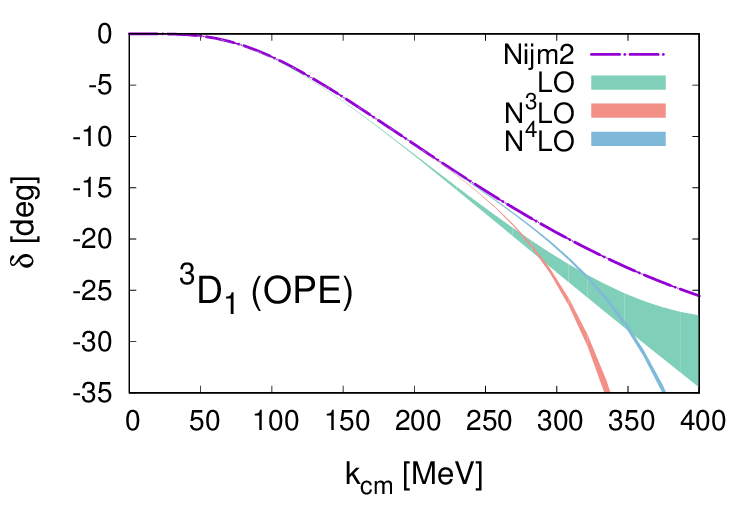}
  \includegraphics[width=4.90cm]{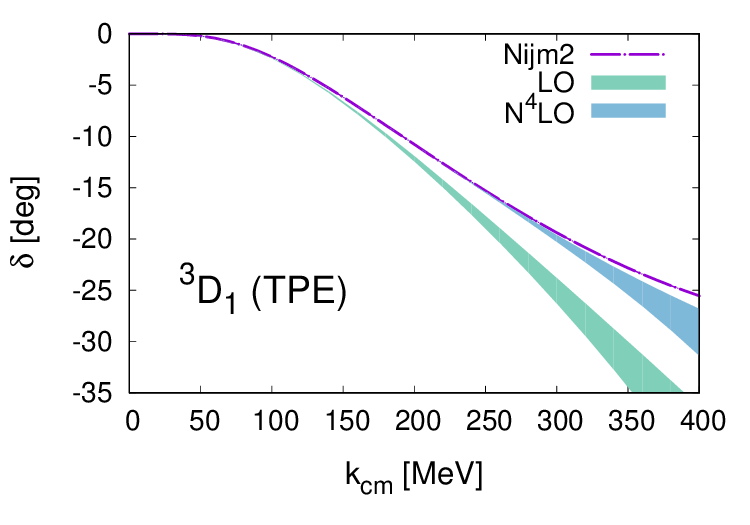}
  \includegraphics[width=4.90cm]{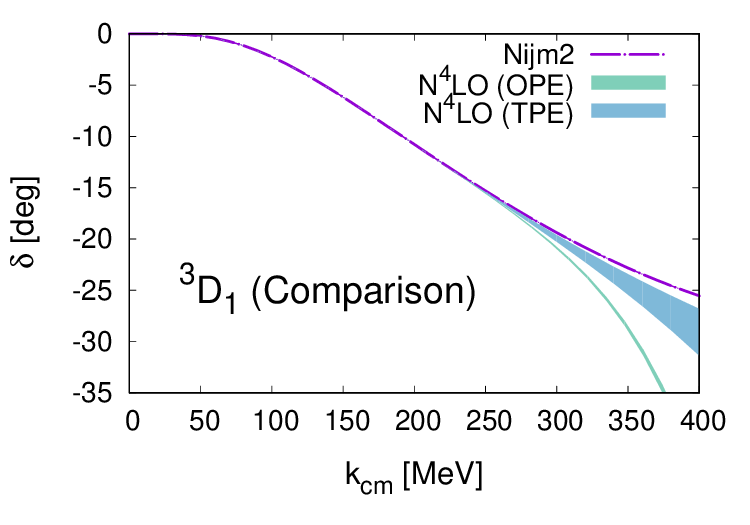}
\end{center}
\caption{Phase shifts of the $^1S_0$, $^3P_0$ and ${}^3S_1$-${}^3D_1$ (nuclear
  bar) partial waves in the two EFT expansions considered in this work,
  EFT(OPE) and EFT(TPE).
  EFT(OPE) is an expansion in which OPE is a non-perturbative ${\rm LO}$ effect
  and happens to be similar to the power counting described
  in Refs.~\cite{Valderrama:2009ei,Valderrama:2011mv},
  except the ${}^1S_0$ partial wave in which EFT(OPE) includes all
  the necessary iterations of $C_2$, which enters at order $Q^0$
  in said expansion.
  EFT(TPE) is an expansion in which TPE is treated non-perturbatively,
  comprising the ${\rm LO}$ in the expansion (though for simplicity
  we also include OPE as part of the ${\rm LO}$ calculation as this
  will not significantly change the results).
  We follow the power counting described in Table~\ref{tab:counting}, where
  only the non-trivial orders (i.e. orders at which a new correction
  is included) are shown.
  The bands correspond to varying the cutoff in the $R_c = (0.5-1.0)\,{\rm fm}$
  range.
  }
\label{fig:comparison}
\end{figure*}

The results are shown in Fig.~\ref{fig:comparison} for the ${}^1S_0$, ${}^3P_0$
and ${}^3S_1$-${}^3D_1$ partial waves, which uses the cutoff range
$R_c = (0.5-1.0)\,{\rm fm}$ to generate bands.
The primary purpose of these bands is to illustrate the size of
the cutoff dependence in the two expansions, while the secondary
purpose is a rudimentary (but cheap) estimation of
the EFT uncertainties (in general, for a properly renormalized theory,
we expect cutoff variations to underestimate these uncertainties,
particularly in the $L \geq 1$ waves; more sophisticated methods to
estimate EFT uncertainties can be consulted
in Refs.~\cite{Schindler:2008fh,Wesolowski:2015fqa,Furnstahl:2015rha,Melendez:2017phj,Wesolowski:2018lzj,Melendez:2019izc,Griesshammer:2015osb,Griesshammer:2020fwr}).
The two expansions work relatively well for most partial waves,
though EFT(TPE) works slightly better (particularly at higher momenta).
Yet, the advantage of EFT(TPE) seems to concentrate mostly in the ${}^1S_0$
partial wave, for which it requires only two couplings (instead of four)
at ${\rm N^4LO}$, with which it also displays a slightly better
agreement with the Nijmegen II pseudodata.
For ${}^3S_1$-${}^3D_1$ we have chosen a counting choice in EFT(OPE) with more
parameters than strictly necessary, just to make calculations simpler.
Thus the comparison made here does not take into account that EFT(OPE) can
be improved as to require less couplings in the deuteron channel.

  In principle it is possible to gain a more detailed understanding of
  the convergence properties of the two expansions from the scaling
  properties of the subleading order corrections, which we analyze
  in Appendix \ref{app:scaling}.
  This suggest that EFT(TPE) might be more convergent than ETF(OPE) not
  only in the $^1S_0$ but also in the $^3S_1$ and $^3P_0$ partial waves.
  Yet, the conclusions are not clear-cut and at most tentative: on the one hand,
  it is not trivial to find dimensionless quantities whose size
  can be conclusively shown to be of ${\mathcal O}(1)$.
  On the other, the simplifying
  counting choices that we have made --- the aforementioned number of
  contact-range couplings in ${}^3S_1$-${}^3D_1$ case, or the fact
  that we promote all the potential to ${\rm LO}$ in EFT(TPE) ---
  obscure the comparisons among the expansions. 

  It should also be noted that we have not tried to make a state-of-the-art
  analysis of the convergence of the EFT(OPE) and EFT(TPE) expansions.
  We merely point out that this could be achieved for instance
  by adapting the Bayesian
  analyses of Refs.~\cite{Schindler:2008fh,Wesolowski:2015fqa,Furnstahl:2015rha,Melendez:2017phj,Wesolowski:2018lzj,Melendez:2019izc}
  to the distorted wave perturbative methods we use here for calculating
  the phase shifts.
  While Bayesian analyses usually deal with non-perturbative scattering
  observables as calculated in the Weinberg prescription (where they
  usually explore the limited range of cutoffs for which this
  prescription works), nothing prevents their application to
  the phase shifts within the two expansions we present here.
  Indeed, Ref.~\cite{Thim:2023fnl} represents a step in this direction,
  though limited to the ${\rm LO}$ of EFT(OPE).
  A potentially interesting advantage of expanding the amplitudes (over
  calculating them non-perturbatively) is the previously mentioned
  determination of the scaling properties of the subleading
  corrections in Appendix \ref{app:scaling}. 
  In this regard, we briefly discuss the possible application of these
  findings for a prospective Bayesian analysis of EFT(OPE) and EFT(TPE)
  in Appendix \ref{app:bayesian}.

\section{Discussion}
\label{sec:discussion}

Here we have considered two possible EFT expansions of the nuclear force:
a typical EFT expansion in which the ${\rm LO}$ is defined by the iteration
of the OPE potential and an atypical one in which TPE is also iterated.
In both cases contact-range operators are included as to guarantee RG
invariance at each order, though a few liberties are taken to ease
the computational burden (particularly in the ${}^3S_1$-${}^3D_1$
channel).
The two expansions, which we have named EFT(OPE) and EFT(TPE), converge
relatively well, though the atypical EFT(TPE) does a better job
in the $^1S_0$ partial wave.

This begs the question: why is this the case? EFT(TPE) is in a sense a
parody of what power counting should be, where contributions that
should enter at $Q^3$ are promoted to $Q^{-1}$.
In the absence of a good physical reason, EFT(TPE) should not
be considered as a legitimate EFT expansion.
The most apparent rationale of why EFT(TPE) might make sense
lies in the slow convergence of the EFT expansion for nuclear physics,
i.e. the poor separation of scales, which makes it plausible
that higher order contributions might accidentally
behave as lower order ones (check the discussion around
Eq.~(\ref{eq:expansion}) for further details).
A cursory look at the S-wave potentials in the chiral limit,
Eqs.~(\ref{eq:V-chiral}-\ref{eq:LambdaTPE-4}), indicates
that parts of TPE are indeed unexpectedly large.

In fact in the chiral limit tensor OPE, TPE(L) and TPE(SL) behave as pure
$1/r^3$, $1/r^5$ and $1/r^6$ power-law infinite-range potentials
(while spin-spin OPE vanishes).
This was beautifully exploited by Birse~\cite{Birse:2005um} to determine
the momenta below which tensor OPE can be treated perturbatively
by applying previously known results from atomic physics
(in particular, the failure of the secular perturbative
expansion~\cite{Cavagnero:1994zz} for the $1/r^3$ potential,
as calculated by Gao~\cite{Gao:1999xyz}).
This suggested that tensor OPE is only perturbative for $k < 66\,{\rm MeV}$
($k < 182\,{\rm MeV}$) in the ${}^3S_1$-${}^3D_1$ ($^3P_0$) partial waves,
at least in the chiral limit, explaining the previous observation that
the Kaplan, Savage and Wise (KSW) counting~\cite{Kaplan:1998tg,Kaplan:1998we}
(which treats all pion exchanges as perturbations) converges slowly and only
for low momenta in the two-nucleon system~\cite{Fleming:1999ee} (check also
Ref.~\cite{Kaplan:2019znu} for a further confirmation of
the limits of perturbative OPE).

It happens that Gao also analyzed the secular perturbative expansion of
attractive $1/r^6$ interactions in Ref.~\cite{Gao:1998zza}, from which
repeating Birse's arguments one obtains a critical momentum
$k_{\rm crit} \simeq (150-200)\,{\rm MeV}$ above which the perturbative
treatment of subleading TPE will not converge
in the chiral limit for the singlet.
Taking into account that in the chiral limit the OPE vanishes in the singlet,
it is sensible to assume that subleading TPE requires a non-perturbative
treatment when $m_{\pi} \to 0$.
In the real world significant deviations should be expected due to
the finite pion mass effects and the OPE ${\rm LO}$ distortion.
Indeed, the EFT(OPE) expansion shows good convergence properties
up to ${\rm N^4LO}$, with no evident signs of a failure
for $k > k_{\rm crit}$.
However, the same was also true for the KSW expansion at ${\rm NLO}$
in the triplets, which later were shown to fail once OPE was iterated
for the first time at ${\rm N^2LO}$~\cite{Fleming:1999ee}.
Thus it might very well happen that the same fate awaits to the EFT(OPE)
expansion once the first iteration of TPE(SL) enters at $Q^7$ /
${\rm N^8LO}$.

For triplet waves the situation is different,
as TPE is now distorted by tensor OPE.
From this, it might be perfectly possible that its strength is screened
by the longer-range OPE distortion, which is why EFT(OPE) works better
in the triplet when compared to the singlet.
Actually, the results we show indicate that there is not a marked difference
between the two expansions for the triplets.
It should also be stressed that for EFT(TPE) we have made the simplification of
promoting the full potential (including OPE) to LO.
It might happen that leaving OPE as a ${\rm NLO}$ contribution, as its NDA
scaling would suggest, might not lead to a converging expansion.
If this were to be the case, the conclusion would be that EFT(TPE) is not
a suitable expansion for the triplets.
From an orthodox EFT perspective this would be good news, as this will
reinforce the standard counting in which OPE drives the low energy
physics of the deuteron, for instance.
But to really confirm this hypothesis we would need to iterate tensor
OPE up to relatively high orders in distorted wave perturbation theory,
which is not exactly easy to do (particularly if we want to guarantee
renormalizability at every step in the calculations).
This would be nonetheless worth exploring in the future.

To summarize, the features of EFT(TPE) are intriguing: it is a really
counterintuitive way of organizing nuclear EFT, yet it provides an
expansion that converges better than the standard way of
organizing the EFT expansion.
One might argue that EFT(TPE) is an ersatz Weinberg prescription,
but it is conceptually different: even though in both cases
we are iterating a large chunk of the effective
finite-range potential, this is not true for the contact-range operators,
most of which are treated as perturbations.
This detail makes EFT(TPE) renormalizable --- the counting of the contacts
is derived from RGE and the amplitudes have a well-defined $R_c \to 0$
limit --- while the Weinberg prescription does not always behave well
in the hard cutoff limit (e.g. the $^1S_0$ channel is not
renormalizable with TPE and two contacts, $C_0$ and $C_2$~\cite{Entem:2007jg},
unless a very specific representation of the contact-range interaction
is invoked~\cite{PavonValderrama:2007nu}).
If anything, the EFT(TPE) expansion provides more questions than answers
and will require further theoretical effort to find its place
within nuclear EFT.

\section*{Acknowledgments}

I would like to thank Ubirajara van Kolck and Michael C. Birse for estimulating
discussions on the topic of this manuscript, Mao-Jun Yan for a careful
reading of this manuscript and the IJCLab of Orsay, where part of
this work was done, for its hospitality.
This work is partly supported by the National Natural Science Foundation
of China under Grants No. 11735003 and No. 11975041, the Fundamental
Research Funds for the Central Universities and the Thousand
Talents Plan for Young Professionals.


\appendix
\section{Expansion of the phase shifts}
\label{app:expansion}

Here we briefly explain how to calculate the perturbative expansion of
the phase shifts in r-space.
The starting point is the reduced Schr\"odinger equation, which reads
\begin{eqnarray}
  -u_k''(r) + \left[ 2\mu V(r) + \frac{l(l+1)}{r^2} \right]\,u_k(r)
  &=& k^2\,u_k(r) \, , \nonumber \\
\end{eqnarray}
with $r$ the radius, $k$ the center-of-mass momentum, $\mu$ the reduced mass of
the system, $l$ the orbital angular momentum, $V$ the two-body potential
and $u_k$ the reduced wave function.
The phase shift can be extracted from the asymptotic behavior of
the reduced wave function at $r \to \infty$
\begin{eqnarray}
  u_k(r) \to {\cot}\delta(k)\,\hat{j}_l(kr) - \hat{y}_l(k r) \, ,
  \label{eq:uk}
\end{eqnarray}
with $\hat{j}_l(x) = x\,j_l(x)$ and $\hat{y}_l(x) = x\,y_l(x)$,
where $j_l(x)$ and $y_l(x)$ are the standard spherical
Bessel functions.

If we now modify the two-body potential as follows
\begin{eqnarray}
  V &\to& V + \delta V \, , 
\end{eqnarray}
this will entail a similar change in the reduced wave function and
the cotangent of the phase shift:
\begin{eqnarray}
  u_k &\to& u_k + \delta u_k \, , \\
  {\rm cot}\delta &\to& {\rm cot}\delta + \delta({\rm cot}\delta) \, . 
\end{eqnarray}
The exact form of the previous modifications can be obtained
from a Wronskian identity involving the reduced Schr\"odinger equations
for $V$ and $V + \delta V$, resulting in the formula
\begin{eqnarray}
  \delta({\rm cot}\delta) = \frac{2\mu}{k}\,
  \int_0^{\infty} dr\,u_k(r)\,\delta V(r)\,\left( u_k(r) + \delta u_k(r) \right)
  \, , \nonumber \\
  \label{eq:delta-cotd}
\end{eqnarray}
where it should be noted that this expression is exact, as we have not
expanded yet in terms of power counting
(or any other expansion scheme).

\subsection{EFT expansion}

For obtaining the perturbative or, more properly, the power counting expansion
of ${\cot} \delta$ we have to consider the explicit expansion of the three
quantities involved
\begin{eqnarray}
  V = \sum_{\nu = \nu_{\rm min}}^{\infty} V^{(\nu)} \, , \\
  u_k = \sum_{\nu = \nu_{\rm min} + 1}^{\infty} u_k^{(\nu)} \, , \\
  \cot{\delta} = \sum_{\nu = \nu_{\rm min} + 1}^{\infty} {[ \cot{\delta} ]}^{(\nu)} \, ,
\end{eqnarray}
and then rearrange the different terms of Eq.~(\ref{eq:delta-cotd}) accordingly,
where care should be taken that only terms of the same order
in the EFT expansion are matched.
This last condition is not trivial at first sight (particularly because
the power counting index of the potential is arranged according to
its p-space representation, see the discussion
around Eq.~(\ref{eq:V-counting}),
yet we are working in r-space),
but can be automated by writing down explicitly
a dummy expansion parameter $\lambda$
\begin{eqnarray}
  V = \lambda^3\,\sum_{\nu = \nu_{\rm min}}^{\infty}\,
  \lambda^{\nu} V^{(\nu)} \, , \\
  u_k = \sum_{\nu = \nu_{\rm min} + 1}^{\infty}  \lambda^{\nu}\,u_k^{(\nu)} \, , \\
  \cot{\delta} = \sum_{\nu = \nu_{\rm min} + 1}^{\infty}\,
  \lambda^{\nu}\,{[ \cot{\delta} ]}^{(\nu)} \, ,
\end{eqnarray}
to which we add these two substitution rules
\begin{eqnarray}
  r \to \frac{r}{\lambda} \quad \mbox{and} \quad k \to \lambda k \, ,
\end{eqnarray}
to account for their power counting.
With this we can readily organize the EFT expansion of the phase shifts
by expanding and rearranging in powers of $\lambda$, and later resetting
this dummy parameter to $\lambda = 1$.

After this, the calculation of the $\nu$-th order phase shifts can be arranged
by means of an iterative process, where the starting point is
the LO phase shift $\delta^{(0)}$, which is calculated
from the reduced Schr\"odinger equation and
the LO potential by standard means.
The NLO phase shift $\delta^{(1)}$ will be obtained
by expanding Eq.~(\ref{eq:delta-cotd}) with the power counting rules
implicitly defined with the dummy $\lambda$ expansion parameter procedure,
leading to
\begin{eqnarray}
  {[ \cot \delta ]}^{(1)}  &=&
  \frac{2\mu}{k} \int_0^{\infty} dr \,u_k^{(0)}(r)\,V^{(0)}(r)\,u_k^{(0)}(r) 
  \nonumber \\
  && \left( = - \frac{\delta^{(1)}}{\sin^2 \delta^{(0)}} \right) \, ,
\end{eqnarray}
where in the second line we explicitly expand ${\cot \delta}^{(1)}$ in terms of
the expansion of the phase shift, i.e. $\delta = \sum_{\nu} \delta^{(\nu)}$.
For ${\rm N^2LO}$, we first have to construct the ${\rm NLO}$ wave function,
which behaves asymptotically as
\begin{eqnarray}
  u_k^{(1)}(r) \to [\cot{\delta}]^{(1)}\,\hat{j}_l(kr) \, ,
\end{eqnarray}
and which we later integrate by means of
\begin{eqnarray}
  -{u_k^{(1)}}''(r) + \left[ 2\mu V^{(-1)}(r) +
    \frac{l(l+1)}{r^2} - k^2\right]\,u_k^{(1)}(r) &=& \nonumber \\
  - 2\mu V^{(0)}(r)\,u_k^{(0)}(r) \, . &&    \nonumber \\
\end{eqnarray}
Once the ${\rm NLO}$ wave function has been computed, the ${\rm N^2LO}$
phase shifts will be given by
\begin{eqnarray}
  {[\cot \delta]}^{(2)}  &=&
  \frac{2\mu}{k} \int_0^{\infty} dr \, u_k^{(0)}(r) \nonumber \\
  && \quad \times \left( V^{(1)}(r) u_k^{(0)}(r) + V^{(0)}(r) u_k^{(1)}(r) \right) 
  \nonumber \\
  && \left( = - \frac{\delta^{(2)}}{\sin^2 \delta^{(0)}} +
  {\left( \frac{\delta^{(1)}}{\sin \delta^{(0)}} \right)}^2\,[\cot \delta]^{(0)} \right) \, ,  \nonumber \\
\end{eqnarray}
where the last line contains the expansion of ${[\cot \delta]}^{(2)}$
in terms of the expansion of the phase shift.
As can be appreciated the expressions become more complex
as the EFT order increases.
Yet, their derivation is straightforward.

\subsection{Regularization and renormalization}

The EFT potential contains a finite- and contact-range piece
\begin{eqnarray}
  V = V_F + V_C \, ,
\end{eqnarray}
with $V_F$ given by pion exchanges and $V_C$ representing the shorter range
physics not explicitly included in the EFT description.
These contributions are singular ($V_F$ involves inverse power-law potentials
and $V_C$ Dirac-delta contributions and their derivatives) and
have to be regularized.
For the finite-range potential we choose a sharp cutoff in r-space, i.e.
\begin{eqnarray}
  V_F(r ; R_c) = V_F(r)\,\theta(r - R_c) \, , \label{eq:sharp-cutoff}
\end{eqnarray}
where $V_F(r; R_c)$ and $V_F(r)$ refer to the regularized and unregularized
version of the finite-range potential.
For the contact-range potential, we choose an energy-dependent
delta-shell regularization
\begin{eqnarray}
  V_C(r; R_c) = \sum_{n=0}^{\infty}
  \frac{C_{2n}(R_c) k^{2n}}{4 \pi R_c^2}\,\delta (r-R_c) \, ,
\end{eqnarray}
where the couplings $C_{2n}(R_c)$ admit a further expansion in terms of
the power counting, i.e. $C_{2n} = \sum_{\nu} C_{2n}^{(\nu)}$, which
we do not explicitly write here.
Finally, for the renormalization of the perturbative phase shifts
we refer to~\cite{Valderrama:2009ei,Valderrama:2011mv}, where
it is explained in detail how the divergences generated
by the finite-range potential are absorbed
by the contact-range couplings.

\subsection{Extension to coupled channels}

For the coupled channels we will adapt the formalism contained
in the Appendix of Ref.~\cite{PavonValderrama:2009nn} and
write the reduced Schr\"odinger equation as
\begin{eqnarray}
  -{\bf u}_k(r) + \left[ 2\mu \,{\bf V}(r) + \frac{{\bf L}^2}{r^2}
    \right]\,{\bf u}_k(r) = k^2\,{\bf u}_k(r) \, , \nonumber \\
\end{eqnarray}
where ${\bf u}_k$ and ${\bf V}$ are now $N \times N$ matrices, while
the squared orbital angular momentum operator takes the form of
the diagonal matrix
\begin{eqnarray}
{\bf L}^2 = {\rm diag}\big( l_1(l_1+1), \dots , l_N(l_N+1) \big) \, ,
\end{eqnarray}
with $l_i$ the orbital angular momentum of channel $i=1, \dots, N$.
The asymptotic behavior of the reduced wave function is given by
\begin{eqnarray}
  {\bf u}_k(r) \to {\bf J}_k(r)\,{\bf M}(k) - {\bf Y}_k(r) \, ,
\end{eqnarray}
where ${\bf J}_k(r)$ and ${\bf Y}_k(r)$ are the diagonal matrices
\begin{eqnarray}
  {\bf J}_k(r) &=& {\rm diag}\,
  \big( \hat{j}_{l_1}(kr), \dots, \hat{j}_{l_N}(k r) \big) \, , \\
  {\bf Y}_k(r) &=& {\rm diag}\,
  \big( \hat{y}_{l_1}(kr), \dots, \hat{y}_{l_N}(k r) \big) \, , 
\end{eqnarray}
with $\hat{j}_l(x)$ and $\hat{y}_l(x)$ as defined below Eq.~(\ref{eq:uk}).
${\bf M}(k)$ is basically the coupled channel version of
${\cot} \delta(k)$, which is related to the S-matrix as
\begin{eqnarray}
  {\bf M}(k) = i \frac{{\bf S}(k) + 1}{{\bf S}(k) - 1} \, .
\end{eqnarray}
It is worth noticing that the standard representation for the reduced wave
function would be an $N$-dimensional vector. Yet, there are $N$ linearly
independent solutions to the previous reduced Schr\"odinger equation,
which allows us to group all the solutions in the ${\bf u}_k$ matrix.
Then, every column of the ${\bf u}_k$ matrix corresponds to
a linearly independent reduced wave function.

As before, if we modify the potential then the subsequent modification
of the ${\bf M}(k)$ matrix and the reduced wave function can be found
by constructing a suitable Wronskian identity, leading to
\begin{eqnarray}
  \delta {\bf M}(k) = \frac{2\mu}{k}\,\int_0^{\infty}\,dr\,
          \left( {\bf u}^T_k(r) + \delta {\bf u}^T_k(r) \right)
          \,\delta {\bf V}(r)\,{\bf u}_k \, , \nonumber \\
\end{eqnarray}
where this expression is exact.
By expanding it in terms of power counting we will find
the explicit expressions for the subleading order phase shits.

Finally, it is useful to write down the explicit expressions for ${\bf M}(k)$
in the eigen representation of the phase shifts~\cite{PhysRev.86.399}.
For the eigen phase shifts, the S-matrix takes the form
\begin{eqnarray}
  {\bf S}(k) =
  {\bf R}(\epsilon)
  \begin{pmatrix}
    e^{2 i \delta_{\alpha}} & 0 \\
    0 &  e^{2 i \delta_{\beta}}
  \end{pmatrix} \,
                {\bf R}(-\epsilon) \, ,
\end{eqnarray}
where $\delta_{\alpha}$ and $\delta_{\beta}$ are the two eigen phase shifts and
${\bf R}$ is the rotation matrix
\begin{eqnarray}
  {\bf R}(\epsilon) =
  \begin{pmatrix}
    \cos{\epsilon} & -\sin{\epsilon} \\
    \sin{\epsilon} &  \cos{\epsilon}
  \end{pmatrix} \, ,
\end{eqnarray}
with $\epsilon$ the mixing angle.
This leads to 
\begin{eqnarray}
  {\bf M}(k) =
  {\bf R}(\epsilon)\,
    \begin{pmatrix}
    \cot{\delta_{\alpha}} & 0 \\
    0 &  \cot{\delta_{\beta}}
    \end{pmatrix} \,
    {\bf R}(-\epsilon)\,
    \, , 
\end{eqnarray}
which happens to be a relatively compact expression.
Putting all the pieces together, we find the next exact expression for
the variation of the phase shifts with respect to the potential
\begin{eqnarray}
  && k\,{\bf R}^{-1}(\epsilon + \delta \epsilon)\,\delta {\bf M}(k)\,
  {\bf R}(\epsilon) = \nonumber \\
  && \quad 2\,\mu\,\int_0^{\infty}\,dr\,{\bf R}^{-1}(\epsilon + \delta \epsilon)
  {\bf u}_k^T(r)\,\delta {\bf V}(r)\,{\bf u}_k(r) {\bf R}(\epsilon) \, .
  \nonumber \\
\end{eqnarray}
Here it is interesting to notice that the ${\bf u}_k {\bf R}$ product
actually coincides with the so-called $\alpha$ and $\beta$ scattering
states
\begin{eqnarray}
  {\bf u}_k(r) {\bf R}(\epsilon) =
  \begin{pmatrix}
    u_{\alpha}(r) & u_{\beta}(r) \\
    w_{\alpha}(r) & w_{\beta}(r) \\
  \end{pmatrix} \, ,
\end{eqnarray}
which are characterized by the relatively simple asymptotic ($r \to \infty$)
forms
\begin{eqnarray}
  \begin{pmatrix}
    u_{\alpha}(r) \\
    w_{\alpha}(r) \\
  \end{pmatrix} \to
  \begin{pmatrix}
    \phantom{+}\cos{\epsilon}\,(\cot{\delta_{\alpha}}\,\hat{j}_{l_a}(k r) -
    \hat{j}_{l_a}(k r)) \\
    \phantom{+}\sin{\epsilon}\,(\cot{\delta_{\alpha}}\,\hat{j}_{l_b}(k r) -
    \hat{j}_{l_b}(k r))
  \end{pmatrix} \, , \nonumber \\ \\
    \begin{pmatrix}
    u_{\beta}(r) \\
    w_{\beta}(r) \\
  \end{pmatrix} \to
  \begin{pmatrix}
    -\sin{\epsilon}\,(\cot{\delta_{\beta}}\,\hat{j}_{l_a}(k r) -
    \hat{j}_{l_a}(k r)) \\
    \phantom{+}\cos{\epsilon}\,(\cot{\delta_{\beta}}\,\hat{j}_{l_b}(k r) -
    \hat{j}_{l_b}(k r))
  \end{pmatrix} \, , \nonumber \\
\end{eqnarray}
with $l_a = j-1$ and $l_b = j+1$ for a standard coupled channel
with total angular momentum $j$.

The power counting expansion of the previous expressions is straightforward
but tedious (particularly if higher order perturbation theory is involved).
However, in most practical settings only first order perturbation theory
is required, in which case we will recover the expressions of
Refs.~\cite{Valderrama:2009ei,Valderrama:2011mv}
(modulo normalization and notational conventions):
\begin{eqnarray}
  {[\cot{\delta_{\alpha}}]}^{(1)} &=&
  \frac{2\mu}{k}\,I^{(0)}_{\alpha \alpha}(k) \, , \\
  {[\cot{\delta_{\beta}}]}^{(1)} &=& \frac{2\mu}{k}\,I^{(0)}_{\beta \beta}(k) \, , \\
  \epsilon^{(1)}\,\left(
  \cot{\delta_{\alpha}^{(0)}} - \cot{\delta_{\beta}^{(0)}} \right) &=&
  \frac{2\mu}{k}\,I^{(0)}_{\alpha \beta}(k) \, ,
\end{eqnarray}
with
\begin{eqnarray}
  I_{\gamma \delta}^{(0)}(k) &=& \int_0^{\infty}\,dr\,\Big[
    u_{\gamma}^{(0)}(r) V_{a a}(r) u_{\delta}^{(0)}(r) \nonumber \\
    && + V_{a b}(r) \left(
    u_{\gamma}^{(0)}(r) w_{\delta}^{(0)}(r) +
    w_{\gamma}^{(0)}(r) u_{\delta}^{(0)}(r) 
    \right) \nonumber \\
    && + w_{\gamma}^{(0)}(r)\, V_{b b}\,w_{\delta}^{(0)}(r) \Big] \, .
\end{eqnarray}

In contrast, had we used the nuclear-bar phase shifts~\cite{PhysRev.105.302}
(i.e. $\bar{\delta}_{\alpha}$, $\bar{\delta}_{\beta}$, $\bar{\epsilon}$)
we would have ended up with bewilderingly complex expressions
that are not particularly suited for practical calculations.
Instead, the best strategy for applying the perturbative expansion
for the nuclear-bar representation is to use the well-known
conversion formulas
\begin{eqnarray}
  \delta_{\alpha} + \delta_{\beta} &=& \bar{\delta}_{\alpha} + \bar{\delta}_{\beta}
  \, , \\
  \sin{(\delta_{\alpha} - \delta_{\beta})} &=&
  \frac{\sin{2 \bar{\epsilon}}}{\sin{2 \epsilon}}  \, , \\
  \sin{(\bar{\delta}_{\alpha} - \bar{\delta}_{\beta})} &=&
  \frac{\tan{2 \bar{\epsilon}}}{\tan{2 \epsilon}} \, , 
\end{eqnarray}
and expand them according to the power counting.

\section{Estimating the convergence of the EFT expansion}
\label{app:scaling}

Here we analyze how the perturbative expansion of the phase shifts
behaves in terms of the involved soft and hard scales.
In particular we are interested in dimensionless quantities that appear
in the EFT expansion and whose size is of $\mathcal{O}(1)$
when the breakdown scale $M$ has been correctly identified.
These quantities might in turn be used to estimate
the convergence of the EFT expansion.

\subsection{Scaling properties}
\label{app:analysis}

We begin by considering the expansion of $\cot{\delta}$ and
how it is calculated, where the idea is to understand
its scaling properties in detail.
Trivially, ${[\cot \delta]}^{(\nu)}$ is expected to scale as
\begin{eqnarray}
  \frac{k}{2\mu}\,{[\cot \delta]}^{(\nu)} &=&
  \int_0^{\infty} dr \, u_k^{(0)}(r)\,{\left[ V(r) u_k(r) \right]}^{(\nu-1)}
  \nonumber \\
  &\propto& {\left( \frac{Q}{M} \right)}^{\nu+1} \, ,
  \label{eq:cotd-scaling}
\end{eqnarray}
where we have explicitly written its expression as a matrix element of
the EFT potential.
The notation $[V u]^{(\mu)}$ indicates all combinations of $V$ and $u$
of order $\mu$, i.e.
\begin{eqnarray}
  {\left[ V(r) u_k(r) \right]}^{(\mu)} = \sum_{\mu_1, \mu_2 \geq 0 / \mu_1 + \mu_2 = \mu}
  V^{(\mu_1)}(r) \, u_k^{(\mu_2)}(r) \, . \nonumber \\
  \label{eq:V-u-exp}
\end{eqnarray}
The integral appearing in the calculation of $[\cot \delta]^{(\nu)}$
can be further subdivided into a polynomial and
non-polynomial contributions:
\begin{eqnarray}
    && \int_0^{\infty} dr \, u_k^{(0)}(r)\,{\left[ V(r) u_k(r) \right]}^{(\nu-1)} =
  \nonumber \\
    &&  \qquad      {\left( \frac{Q_{\slashed{k}}}{M}\,\right)}^{\nu + 1}\,
            \left[ {f}^{(\nu)}_{\rm pol}(\frac{k}{Q_{\slashed{k}}})
              + {f}^{(\nu)}_{\rm nonpol}(\frac{k}{Q_{\slashed{k}}}) \right] \, ,
            \label{eq:f-pol-nonpol}
\end{eqnarray}
which we have written as functions of the dimensionless ratio
$k / Q_{\slashed{k}}$, where $Q_{\slashed{k}}$ represents all
the light scales with the exception of the momentum $k$.
Here by polynomial we specifically mean the part of the integral that
behaves as a pure polynomial on the momentum $k$ regardless of
how large $k$ is.
The non-polynomial part will still be expansible in powers of $k$ at small
momenta, but this expansion is not expected to converge
for momenta $k \gg m_{\pi}$.
We will study each contribution in more detail below,
including how this division into polynomial and
non-polynomial pieces appears.

The polynomial part stems from (i) the contact-range potential and (ii)
the divergences of the finite-range potential (which after
renormalization are absorbed in the contact-range couplings).
If we regularize with a sharp cutoff in r-space
(i.e. as in Eq.~\ref{eq:sharp-cutoff}),
we may write these two contributions separately as:
\begin{eqnarray}
  && \int_0^{\infty} dr \, u_k^{(0)}(r)\,{\left[ V_C(r) u_k(r) \right]}^{(\nu-1)} =
  \nonumber \\
  && \quad
  \sum_{n=0}^{n_{\rm max}(\nu)} \frac{u_k^{(0)}(R_c) \left[ C_{2n}(R_c) u_k(R_c) \right]^{(\nu-1)}}{4 \pi R_c^2}\,k^{2n}\, , \\
    && \int_0^{\infty} dr \, u_k^{(0)}(r)\,{\left[ V_F(r) u_k(r) \right]}^{(\nu-1)} =
  \nonumber \\
  && \quad
  {\left(\frac{Q_{\slashed{k}}}{M}\right)}^{\nu}\,
  {\left(\frac{Q_{\slashed{k}}}{k}\right)}^{2l}\,
  \sum_{n=0}^{n_{\rm max}(\nu)}\,
  {d}_{2n}\,{\left(\frac{k}{Q_{\slashed{k}}}\right)}^{2n}\,
  \times \nonumber \\ && \qquad
  \int_{(Q_{\slashed{k}}\,R_c)} \frac{dx}{x^{3+\nu-2\alpha-2n}} 
  \,
  + \mbox{(finite terms)} \, , \label{eq:VF-integral}
\end{eqnarray}
where $V_C$ and $V_F$ are the contact- and finite-range pieces of
the EFT potential and the notation $[ \dots ]^{(\mu)}$ was previously
defined in Eq.~(\ref{eq:V-u-exp}).
Here $l$ is the orbital angular momentum, $\alpha$ refers to the power-law
behavior of the ${\rm LO}$ wave function at short distances (which might
depend on $l$ if the ${\rm LO}$ potential is regular), $n_{\rm max}(\nu)$
the number of contact-range couplings at $\nu$-th order and
${d}_{2n}$ is a dimensionless coefficient that
characterizes a given divergence.
The notation
\begin{eqnarray}
  \int_{(Q_{\slashed{k}}\,R_c)} \frac{dx}{x^{3+\nu-2\alpha-2n}} \, ,
\end{eqnarray}
where only the lower integration bound is shown, has been used to indicate
that we are interested in the divergent part of this integral.
If we now combine the two terms together, after removing the divergences
we will find the polynomial part of the perturbative integral:
\begin{eqnarray}
  && \int_0^{\infty} dr \, u_k^{(0)}(r)\,{\left[ V(r) u_k(r) \right]}^{(\nu-1)}
  \, \Big|_{\rm pol} =
  \nonumber \\
  && \quad
  {\left(\frac{Q_{\slashed{k}}}{M}\right)}^{\nu}\,
  {\left(\frac{Q_{\slashed{k}}}{k}\right)}^{2l}\,
  \sum_{n=0}^{n_{\rm max}(\nu)}\,{c}_{2n}\,{\left(\frac{k}{Q_{\slashed{k}}}\right)}^{2n}
  \nonumber \\
  && \qquad + \quad \mbox{(cut-off dependent polynomials)} \, ,
\end{eqnarray}
where there is a cut-off dependent part of the polynomial (comprising higher
powers of $k$), which vanishes when $R_c \to 0$.
Even though derived from a particular regulator, we expect the result
above to be independent of the choice of regulator.

The non-polynomial piece stems from the finite terms of
Eq.~(\ref{eq:VF-integral}) and can in fact be shown to be non-polynomial
if we consider that --- for $r \to \infty$ --- we have on the one hand
that the finite-range potential decays exponentially
\begin{eqnarray}
  V_F^{(\nu)}(r) \to \frac{1}{(M r)^{\nu +2}}\,P_F(Q_{\slashed{k}} r)\,
  \frac{e^{-2 m r}}{r} \, ,
\end{eqnarray}
where $P_F(x)$ represents a polynomial, and, on the other,
that the wave function can also be written in terms
of exponentials
\begin{eqnarray}
  u_k(r) \to e^{i \delta_l(k)}\,e^{i (k r - l \pi/2)} -
  e^{-i \delta_l(k)}\,e^{-i (k r - l \pi/2)} \, .
  \nonumber \\
\end{eqnarray}
Combining these two observations it can be appreciated that
the infrared behavior of the integral
\begin{eqnarray}
  && \int^{\infty} dr \, u_k^{(0)}(r)\,{\left[ V_F(r) u_k(r) \right]}^{(\nu-1)}
  \propto
  \nonumber \\ &&
  \frac{e^{\pm 2 i \delta_l^{(0)}(k)}}{M^{\nu+2}}\,\int^{\infty} dr \, P_F(Q_{\slashed{k}} r)\,\frac{e^{(\pm 2 i k -2 m) r}}{r^{\nu+3}} + \dots \, , \nonumber \\
\end{eqnarray}
diverges for $|{\rm Im}\,k| > m$, giving rise to two branch cuts
beginning at $k = \pm i m$.
As a consequence, the non-polynomial part of the integral does not have
a convergent Taylor expansion in terms of the momenta for $k > m$.

Here we notice that we have explicitly assumed that the subleading potential is
dominated by two-pion exchanges: hence the $e^{-2 m r}$ decay.
This is not strictly true though, as there are subleading corrections to
the OPE potential too.
These corrections, if treated perturbatively, will generate additional
branch cuts at $k = \pm i m/2$.
Yet, the bottom-line is the same: the non-polynomial part is in fact
non-polynomial for large enough momenta.

\subsection{Naturalness}

Ideally, we want to extract coefficients in the EFT expansion
whose size is $\mathcal{O}(1)$.
The most direct candidate would be the dimensionless function ${f}^{(\nu)}(x)$
that appears when rescaling $[\cot \delta]^{(\nu)}$
according to its power counting
\begin{eqnarray}
  {\left( \frac{M}{Q_{\slashed{k}}}\,\right)}^{\nu + 1}\,\frac{k}{2\mu}\,[\cot \delta]^{(\nu)} = f^{(\nu)}(\frac{k}{Q_{\slashed{k}}}) \, , \label{eq:f-def}
\end{eqnarray}
but there is the problem that $\cot{\delta}$ diverges when $\delta \to 0$.
This makes it impractical to argue whether ${f}^{(\nu)}$
is $\mathcal{O}(1)$ or not.

This limitation is easily circumvented by rescaling the phase shift
$\delta^{(\nu)}$ instead, resulting in the definition of
\begin{eqnarray}
  {\left( \frac{M}{Q_{\slashed{k}}}\,\right)}^{\nu + 1}\,\frac{k}{2\mu}\,\delta^{(\nu)} &=& {g}^{(\nu)}(\frac{k}{Q_{\slashed{k}}}) \, ,
\end{eqnarray}
which is now finite when $\delta^{(0)} \to 0$ and related to $f^{(\nu)}$ by
\begin{eqnarray}
  {g}^{(\nu)} &=& -\sin^2{\delta^{(0)}}\,{f}^{(\nu)} + \dots
  \dots \, , \label{eq:g-reexpansion}
\end{eqnarray}
where the dots represent further terms stemming from the expansion of
$[\cot \delta]^{(\nu)}$, i.e. products of $f^{(\nu_1)} f^{(\nu_2)}$ such
that $\nu_1 + \nu_2 = \nu$.

The dimensionless function ${g}^{(\nu)}(x)$ is not expected to
diverge at normal momenta (as it is derived from phase shifts,
which do not diverge), and is thus a better choice
for naturalness.
Its size is not necessarily $\mathcal{O}(1)$, but this is not crucial though
if our intention is to compare the convergence of
the EFT(OPE) and EFT(TPE) expansions.
In fact we do not need to know the hard scale $M$ of the EFT
we are considering, but just compare both expansions to a common scale
\begin{eqnarray}
  {\left( \frac{M'}{\Lambda_{\rm OPE}}\,\right)}^{\nu + 1}\,\frac{k}{2\mu}\,\delta^{(\nu)} &=& {\left( \frac{M'}{M_{\rm OPE}} \right)}^{\nu+1} {g}^{(\nu)}_{\rm OPE}(\frac{k}{\Lambda_{\rm OPE}})  \nonumber \\ 
  &=& \hat{g}^{(\nu)}_{\rm OPE}(\frac{k}{\Lambda_{\rm OPE}}) \, , \label{eq:gOPE} \\
  {\left( \frac{M'}{\Lambda_{\rm TPE}}\,\right)}^{\nu + 1}\,\frac{k}{2\mu}\,\delta^{(\nu)} &=& {\left( \frac{M'}{M_{\rm TPE}} \right)}^{\nu+1} {g}^{(\nu)}_{\rm TPE}(\frac{k}{\Lambda_{\rm TPE}}) \nonumber \\
  &=& \hat{g}^{(\nu)}_{\rm TPE}(\frac{k}{\Lambda_{\rm TPE}}) \, , \label{eq:gTPE}
\end{eqnarray}
where $\Lambda_{\rm OPE}$, $M_{\rm OPE}$ and $\Lambda_{\rm TPE}$, $M_{\rm TPE}$ are
the typical light and hard scales for the EFT(OPE) and EFT(TPE)
expansions, and with $M'$ an arbitrary scaling mass.
Regardless of the numerical factor to which the dimensionless functions
${g}$ or $\hat{g}$ gravitate, we nonetheless expect that
\begin{eqnarray}
  \mathcal{O}\,\left( \hat{g}_{\rm TPE} \right) =
  \mathcal{O}\,\left(
          {\left(\frac{M_{\rm OPE}}{M_{\rm TPE}}\right)}^{(\nu+1)} \hat{g}_{\rm OPE} \right) \, . \label{eq:g-breakdown-scaling}
\end{eqnarray}
That is, when comparing the dimensionless coefficients of the two expansions
we do not need to know the specific numerical factors.

\begin{figure}[ttt]
  \begin{center}
  \includegraphics[width=6.25cm]{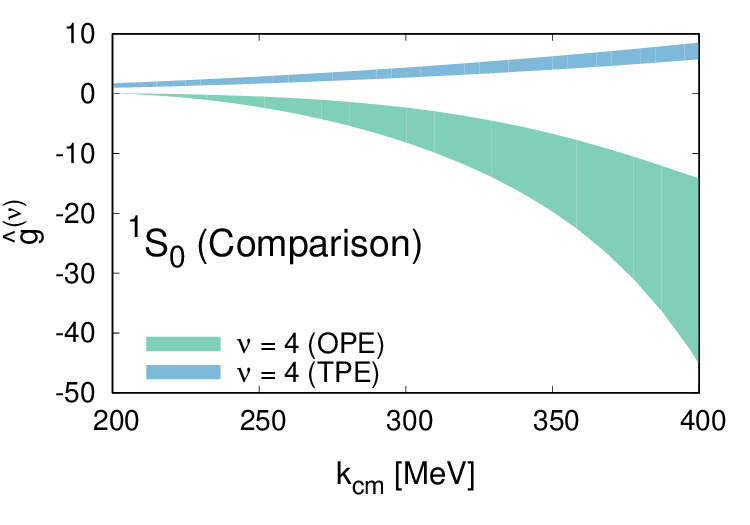}
  \includegraphics[width=6.25cm]{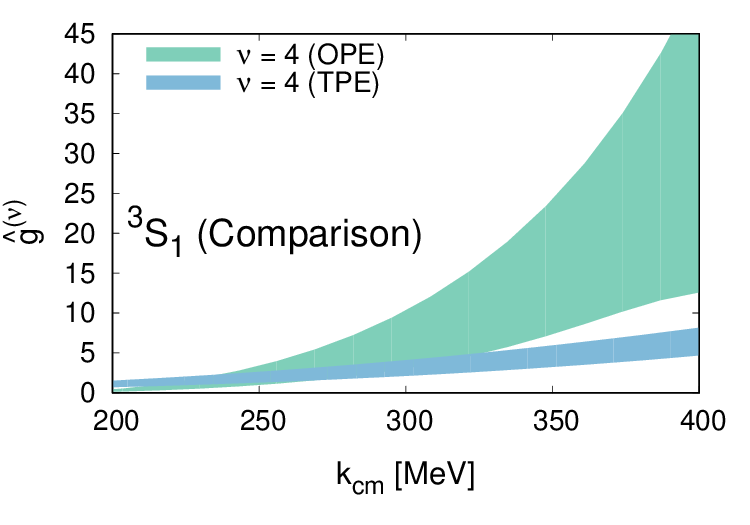}
  \includegraphics[width=6.25cm]{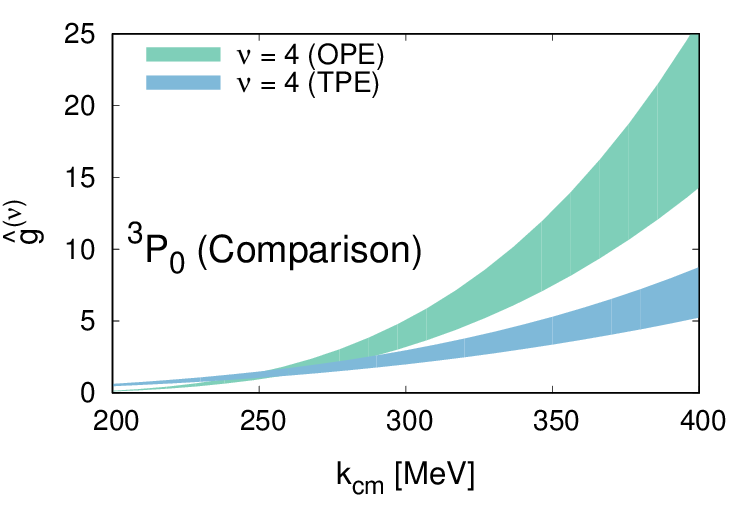}
\end{center}
  \caption{
    The rescaled dimensionless phase shift $\hat{g}^{\nu}$ as defined
    in Eqs.~(\ref{eq:gOPE}) and (\ref{eq:gTPE}) for the EFT(OPE) and
    EFT(TPE) expansions. It can be appreciated that $\hat{g}^{\nu}$
    is smaller for the expansion around TPE than for the OPE one,
    suggesting better convergence.
  }
\label{fig:rescaling-comparison-full}
\end{figure}

With this we now compare the functions $\hat{g}^{(\nu)}$ for $\nu = 4$
in the EFT(OPE) and EFT(TPE) expansions for the $^1S_0$, $^3S_1$ and
$^3P_0$ partial waves.
We take $M' = 0.5\,{\rm GeV}$, which is a typical guess for the hard
scale of the nuclear EFT expansion, $\Lambda_{\rm OPE} = m_{\pi} = 138\,{\rm MeV}$
(i.e. the pion mass) and $\Lambda_{\rm TPE} = 233\,{\rm MeV}$ (i.e. the natural
scale of the subleading TPE potential in the singlet,
see Eq.~(\ref{eq:LambdaTPE-2})).
The comparison is shown in Fig.~\ref{fig:rescaling-comparison-full},
from which it seems that EFT(TPE) is more convergent
than EFT(OPE) for the $^1S_0$, $^3S_1$ and $^3P_0$
partial waves.

However, if we give it a bit of thought, it will become apparent
that the previous comparison might be flawed.
The reason is the polynomial part of these dimensionless function,
which on closer inspection contains positive powers of
$k / Q_{\slashed{k}}$.
If we expand the polynomial in $[\cot \delta]^{(\nu)}$ (and momentarily
ignore cutoff dependent terms)
\begin{eqnarray}
  && {\left( \frac{M}{Q_{\slashed{k}}}\,\right)}^{\nu + 1}\,\frac{k}{2\mu}\,[\cot \delta]^{(\nu)} = {f}^{(\nu)}(\frac{k}{Q_{\slashed{k}}}) \nonumber \\
  && \quad = {f}_{\rm pol}^{(\nu)}(\frac{k}{Q_{\slashed{k}}}) +
  {f}_{\rm nonpol}^{(\nu)}(\frac{k}{Q_{\slashed{k}}}) \nonumber \\
  && \quad = \,{c}_0^{(\nu)}\,{\left( \frac{k}{Q_{\slashed{k}}} \right)}^{-2l}
  + {c}_2^{(\nu)}\,{\left( \frac{k}{Q_{\slashed{k}}} \right)}^{2-2l} + \dots
  \nonumber \\
  && \quad + \,{c}_{2 n_{\rm max}}^{(\nu)}\,{\left( \frac{k}{Q_{\slashed{k}}} \right)}^{2 n_{\rm max}-2l} + {f}_{\rm nonpol}^{(\nu)}(\frac{k}{Q_{\slashed{k}}}) \, ,
  \label{eq:f-expansion}
\end{eqnarray}
we realize that this is indeed a problem in central waves.
For instance, in the $^1S_0$ partial wave, we will have powers of
$(k/m_{\pi})^6$ for EFT(OPE) at $\nu = 3,4$, which will grow
disproportionally large in the $m_{\pi} < k < M$ region.
Notice that there is no difference if we expand $\delta^{(\nu)}$ instead of
$[\cot \delta]^{\nu}$, the problem will still be present either way.

The solution is simple --- to remove the polynomial part coming from the
expansion of $[\cot \delta]^{\nu}$ --- in which case we are left with
\begin{eqnarray}
  && {\left( \frac{M}{Q_{\slashed{k}}}\,\right)}^{\nu + 1}\,\frac{k}{2\mu}\,\delta^{(\nu)} + \sin^2\,\delta^{(0)}\,{f}_{\rm pol}^{(\nu)}(\frac{k}{Q_{\slashed{k}}}) + \dots
  \nonumber \\
  && \quad   = {g}_{\rm nonpol}^{(\nu)}(\frac{k}{Q_{\slashed{k}}}) \, ,
\end{eqnarray}
where ${g}_{\rm nonpol}^{(\nu)}$ is, in a first instance, not expected to grow
as a polynomial, at least in the $m_{\pi} < k < M$ range.
The dots represent further polynomial terms from the iteration of subleading
terms, check Eq.~(\ref{eq:g-reexpansion}) and the discussion
around it.
These terms do only appear in the EFT(OPE) expansion of the $^1S_0$ partial
wave, yet they can be ignored by redefining $f^{(\nu)}_{\rm pol}$ as
the polynomial piece of
\begin{eqnarray}
  - \frac{1}{\sin^2\,\delta^{(0)}} {\left( \frac{M}{Q_{\slashed{k}}}\,\right)}^{\nu + 1}\,\frac{k}{2\mu}\,\delta^{(\nu)} =
  \bar{f}_{\rm pol}^{(\nu)} + \bar{f}_{\rm nonpol}^{(\nu)} \, ,
  \nonumber \\
\end{eqnarray}
instead of its usual definition in terms of $[\cot \delta]^{(\nu)}$.
In most cases this redefinition has no appreciable impact on the numerical
values of the coefficients of the polynomial $f^{(\nu)}_{\rm pol}$ /
$\bar{f}^{(\nu)}_{\rm pol}$, the only exception being $^1S_0$
in EFT(OPE) for $\nu=3,4$ where the relative differences
can reach $(10-30)\%$ for $c_4^{(\nu)}$.

Finally, by rescaling again with an arbitrary mass $M'$ 
\begin{eqnarray}
    && {\left( \frac{M'}{\Lambda_{\rm OPE}}\,\right)}^{\nu + 1}\,\frac{k}{2\mu}\,\delta^{(\nu)} + \sin^2\,\delta^{(0)}\,\hat{f}_{\rm pol,OPE}^{(\nu)}(\frac{k}{\Lambda_{\rm OPE}}) \nonumber \\
  && \quad = \hat{g}_{\rm nonpol,OPE}^{(\nu)}(\frac{k}{\Lambda_{\rm OPE}}) \, , \\
      && {\left( \frac{M'}{\Lambda_{\rm TPE}}\,\right)}^{\nu + 1}\,\frac{k}{2\mu}\,\delta^{(\nu)} + \sin^2\,\delta^{(0)}\,\hat{f}_{\rm pol,TPE}^{(\nu)}(\frac{k}{\Lambda_{\rm TPE}}) \nonumber \\
  && \quad = \hat{g}_{\rm nonpol,TPE}^{(\nu)}(\frac{k}{\Lambda_{\rm TPE}}) \, , 
\end{eqnarray}
we obtain the rescaled $\hat{g}_{\rm nonpol,OPE}^{(\nu)}$ and
$\hat{g}_{\rm nonpol,TPE}^{(\nu)}$, without the unwanted
contamination of the fast-growing polynomial terms.
Notice that $\hat{f}_{\rm pol,OPE}^{(\nu)}$ and $\hat{f}_{\rm pol,TPE}^{(\nu)}$
are just polynomials in $k^2/\Lambda_{\rm OPE}^2$ and $k^2/\Lambda_{\rm TPE}^2$,
which can be directly fitted to the $\delta^{(\nu)}$
phase shifts at low momenta.

\begin{figure}[ttt]
  \begin{center}
  \includegraphics[width=6.25cm]{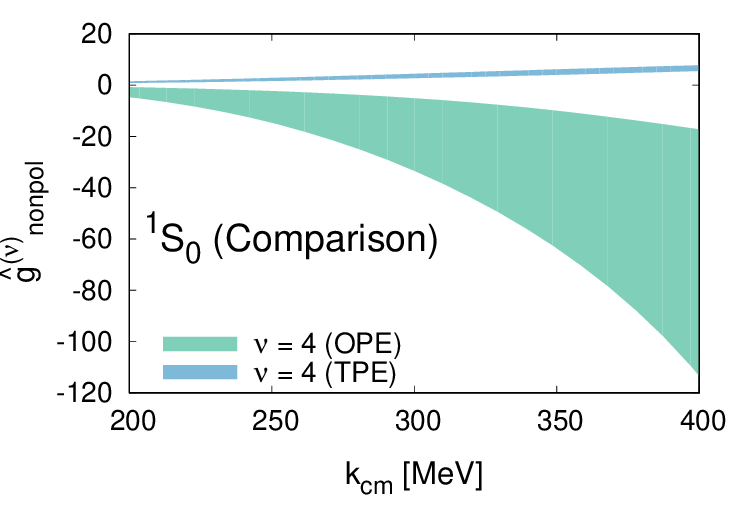}
  \includegraphics[width=6.25cm]{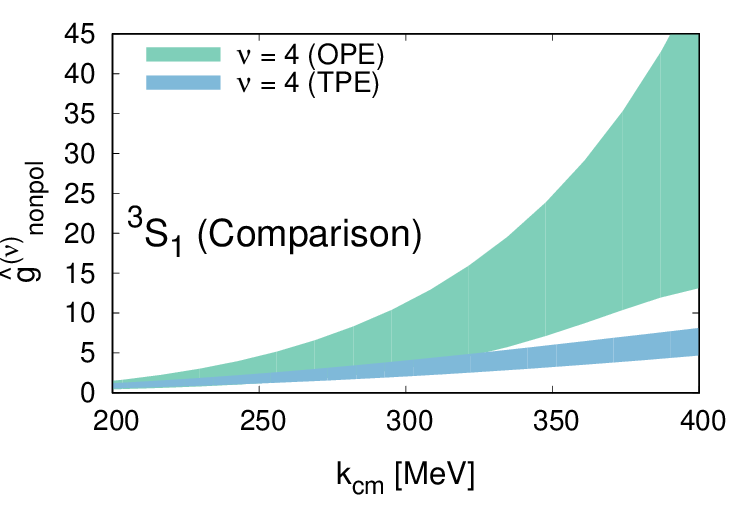}
  \includegraphics[width=6.25cm]{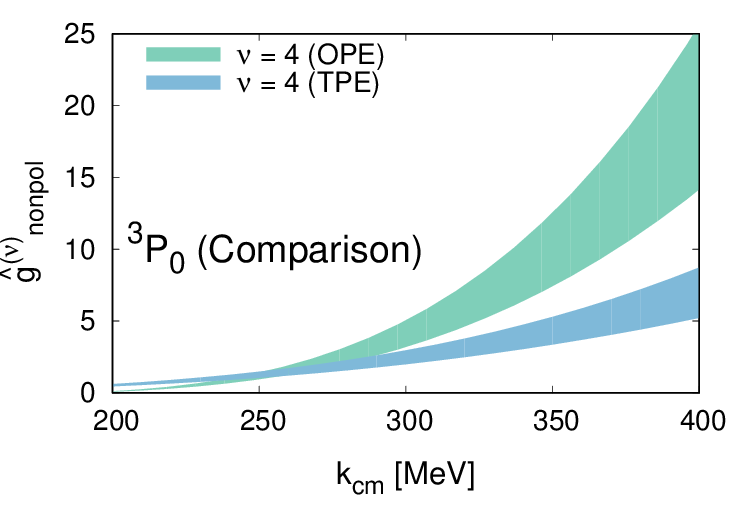}
\end{center}
  \caption{
    The rescaled dimensionless phase shift $\hat{g}^{\nu}_{\rm nonpol}$ once
    the polynomial contributions have been removed, as explained and
    define around Eqs.~(\ref{eq:gOPE}) and (\ref{eq:gTPE}).
    In general there is no appreciable difference with respect to leaving
    the polynomial contributions, indicating that they are masked either
    by the energy dependence stemming from the subleading contributions
    to the finite-range potential in the case of EFT(OPE) or by finite
    cutoff effects.
  }
\label{fig:rescaling-comparison-nonpol}
\end{figure}

Yet, when we compare the new functions $\hat{g}^{(\nu)}_{\rm nonpol}$ for $\nu = 4$
in the EFT(OPE) and EFT(TPE) expansions for the $^1S_0$, $^3S_1$ and
$^3P_0$ partial waves in Fig.~\ref{fig:rescaling-comparison-nonpol},
not much of a difference can be appreciated with respect to
the previous $\hat{g}^{(\nu)}$ that still contained
polynomial contributions.
The exception is the $^1S_0$ partial wave, where removing the polynomial
worsens the comparison for EFT(OPE).
For the new comparison in Fig.~\ref{fig:rescaling-comparison-nonpol},
we have again taken $M' = 0.5\,{\rm GeV}$,
$\Lambda_{\rm OPE} = m_{\pi} = 138\,{\rm MeV}$
and $\Lambda_{\rm TPE} = 233\,{\rm MeV}$. 
The polynomial in $\hat{f}^{(\nu)}$ has been fitted to the $\delta^{(\nu)}$ phase
shifts in the $k = (20-100)\,{\rm MeV}$ region by assuming that at low
momenta
\begin{eqnarray}
  && - \frac{1}{\sin^2\,\delta^{(0)}} {\left( \frac{M}{Q_{\slashed{k}}}\,\right)}^{\nu + 1}\,\frac{k}{2\mu}\,\delta^{(\nu)} \nonumber \\
  && \approx (\frac{Q_{\slashed{k}}}{k})^{2l}\,\left[ \sum_{n=0}^{n_{\rm max}(\nu)} c_{2n}^{(\nu)}\,(\frac{k}{Q_{\slashed{k}}})^{2n} + d_{\rm res}\,\,(\frac{k}{Q_{\slashed{k}}})^{2n_{\rm max} + 2}
    \right] \, , \nonumber \\ \label{eq:f-expansion-effective}
\end{eqnarray}
with $n_{\rm max}(\nu)$ the number of contact interactions at the order
we are considering and where we add an additional order to the polynomial
to stabilize the extracted numerical values of the $c_{2n}^{(\nu)}$ coefficients.
This extra order and its coefficient $d_{\rm res}$ are not
part of $\hat{f}^{(\nu)}$ and are discarded once we fit
the low energy $\delta^{(\nu)}$ phase shifts.
Notice that there is an offset of one order between the potential and
the phase shifts. That is, for a given partial wave, the number of
terms in the $\hat{f}^{(\nu)}$ polynomial is the same as the number
of contact-range couplings in the $Q^{\nu-1}$ potential
(as shown in Table \ref{tab:counting}).

From the previous, the expectation is still that EFT(TPE) is more convergent
than EFT(OPE) for the $^1S_0$, $^3S_1$ and $^3P_0$ partial waves.
The naive application of the scaling of Eq.~(\ref{eq:g-breakdown-scaling})
to $\hat{g}_{\rm nonpol, OPE}$ and $\hat{g}_{\rm nonpol, TPE}$ would suggest
that:
\begin{eqnarray}
  M_{\rm OPE} \sim (0.7-0.8)\,M_{\rm TPE} \, . \label{eq:M-EFT-comparison}
\end{eqnarray}
But caution is advised: the $\nu=4$ contribution to EFT(OPE) contains both
a contact- and finite-range contribution, the later being subleading TPE.
In contrast, the $\nu=4$ contribution to EFT(TPE) comes exclusively from
the subleading contact-range potential: if the cutoff were to be
removed, it should become purely polynomial.
Indeed, the $\hat{g}^{(\nu)}_{\rm nonpol, TPE}$ functions
in Fig.~\ref{fig:rescaling-comparison-nonpol} are finite-cutoff effects
that should vanish as $R_c \to 0$.
Because of this, the comparison of Fig.~\ref{fig:rescaling-comparison-nonpol}
is biased against EFT(OPE).
Thus we either go back to the full $\hat{g}^{(\nu)}_{\rm TPE}$ (with
the polynomials), as already done in Fig.~\ref{fig:rescaling-comparison-full},
or move to $\nu=5$ where both expansion receive new contributions
from the finite-range potential.

Yet, there is the additional issue of the dimensionless function
${g}^{(\nu)}_{\rm nonpol}$ and its expected size.
A more careful analysis of its properties is presented
in Appendix \ref{app:nonpol}, which suggests that its size
is $\mathcal{O}( 2^{\nu}\,\pi / (\nu+1)! )$.
If this is correct, it would imply that the non-polynomial contributions are
irrelevant for the convergence of the EFT expansion (they would always
converge), and hence its size does not necessarily provide useful
information about the breakdown scale.
Besides, the initial assumption we made that the ${g}^{(\nu)}_{\rm nonpol}$ do not
grow as fast as the polynomial part at moderate momenta might also
be flawed, as their behavior grows either as an odd power of $k$
or an even power multiplied by a logarithm (though this is only true
in the $m/k \to 0$ limit and hence might not fully materialize
in the $m_{\pi} < k < M$ regime), see Eqs.~(\ref{eq:f-nu-even})
and (\ref{eq:f-nu-even}).

\begin{table}[thh]
\begin{center}
\begin{tabular}{|c|c|c|}
  \hline
  \noalign{\smallskip}
  \hline
  {\multirow{2}{*}{Coefficient}} & \multicolumn{2}{c|}{${}^1S_0$} \\
  \cline{2-3}
  & EFT(OPE) & EFT(TPE) \\
  \hline
  $c^{(1)}_0$ & $\mathcal{O}(10^{-3})$ & - \\
  $c^{(2)}_0$ & $\mathcal{O}(10^{-3})$ & - \\
  $c^{(3)}_0$ & $-(1.35-0.91) \cdot 10^{-2}$ & - \\
  $c^{(4)}_0$ & $-(2.94-0.48) \cdot 10^{-2}$ & $\mathcal{O}(10^{-2})$ \\
  \hline
  $c^{(1)}_2$ & $+(0.709-0.418)$ & - \\
  $c^{(2)}_2$ & $-(3.32-3.43) \cdot 10^{-2}$ & - \\
  $c^{(3)}_2$ & $+(0.424-0.320)$ & - \\
  $c^{(4)}_2$ & $+(0.780-0.137)$ & $-(0.99-1.81) $ \\
  \hline
  $c^{(3)}_4$ & $-(1.07-0.91)$ & - \\
  $c^{(4)}_4$ & $-(1.53-0.30)$ & -  \\
  \hline
  \noalign{\smallskip}
  \hline
  {\multirow{2}{*}{Coefficient}} & \multicolumn{2}{c|}{${}^3S_1$} \\
  \cline{2-3}
  & EFT(OPE) & EFT(TPE) \\
  \hline
  $c^{(3)}_0$ & $+(0.234-0.205)$ & - \\
  $c^{(4)}_0$ & $-(0.869-0.317)$ & $+(1.25-2.64) \cdot 10^{-2}$ \\
  \hline
  $c^{(3)}_2$ & $(+0.47)-(-1.10)$ & - \\
  $c^{(4)}_2$ & $+(1.26-0.50)$ & $-(0.37-1.02)$ \\
  \hline
  \noalign{\smallskip}
  \hline
  {\multirow{2}{*}{Coefficient}} & \multicolumn{2}{c|}{${}^3P_0$} \\
  \cline{2-3}
  & EFT(OPE) & EFT(TPE) \\
  \hline
  $c^{(3)}_0$ & $-(0.240-0.249)$ & - \\
  $c^{(4)}_0$ & $-(0.317-0.392)$ & $-(3.04-3.26) \cdot 10^{-2}$ \\
  \hline
  $c^{(3)}_2$ & $-(1.46-1.52)$ & - \\
  $c^{(4)}_2$ & $-(1.18-1.52)$ & $-(0.517-0.552)$ \\
  \hline \hline
\end{tabular}
\end{center}
\caption{
  Dimensionless coefficients $c^{(\nu)}_{2n}$ of the low momentum $k^2$
  expansion of the function $f^{(\nu)}$ (basically, the $\nu$-th order
  contribution to the phase shift rescaled according to
  the power counting) as defined in Eq.~(\ref{eq:f-expansion}).
  For their numerical determination, we have fitted ${f}^{(\nu)}$ to
  the form of Eq.~(\ref{eq:f-expansion-effective}) in the
  $k = (20-100)\,{\rm MeV}$ region.
  For the EFT(OPE) expansion we set
  $Q_{\slashed{k}} = m_{\pi} \simeq 138\,{\rm MeV}$, while for EFT(TPE)
  we use $Q_{\slashed{k}} = \Lambda_{\rm TPE}({}^1S_0) \simeq 233\,{\rm MeV}$.
  In both cases we take $M' = 500\,{\rm MeV}$.
  Their values are shown as an interval, with the first (second) number
  corresponding to a cutoff of $R_c = 0.5\,{\rm fm}$ ($1\,{\rm fm}$).
  If no numerical value is above $10^{-2}$ within the interval, we simply
  indicate the order of magnitude of the coefficients.
}
\label{tab:coefficients}
\end{table}

Alternatively, there is the possibility of comparing the $c_{2n}^{(\nu)}$
coefficients comprising the polynomials $\hat{f}^{(\nu)}$
in the two expansions.
From what we have discussed until this point, these coefficients are probably
the best candidate for a dimensionless quantity behaving as $\mathcal{O}(1)$
within the EFT expansion.

In this case we arrive at the coefficients shown in Table \ref{tab:coefficients}
for the $^1S_0$, $^3S_1$ and $^3P_0$ partial waves.
There it can be appreciated that a few coefficients are unnaturally small ---
from $\mathcal{O}(10^{-2})$ to $\mathcal{O}(10^{-3})$ --- , though this is a
consequence of the fact that these coefficients are simply higher order
corrections to information that was already included at lower orders.
For instance, the $c_0^{(\nu)}({}^1S_0)$ coefficients are subleading
contributions to the scattering length $a_0$. Thus, if $a_0$ has been
already fixed at lower orders (which is indeed the case for the singlet),
then the size of the subleading corrections will be considerably smaller
than the naive expectation coming from the power counting.

For coefficients that carry new information, their size tends to be
$\mathcal{O}(1)$, which actually is what we would expect
if we have correctly identified the soft and hard
scales of the expansion.
For the singlet, the $\nu=3,4$ $c_{2}^{(\nu)}$ and $c_{4}^{(\nu)}$ coefficients
are all $\mathcal{O}(1)$, independently of whether we are considering
EFT(OPE) or EFT(TPE), see Table \ref{tab:coefficients}.
In fact it could even be argued that EFT(OPE) might work better than EFT(TPE)
in the singlet.
The situation is the opposite in the $^3S_1$ and $^3P_0$ cases, where
$c_{2}^{(\nu=4)}$ turns out to be a bit smaller in EFT(TPE) than EFT(OPE).

There is however a potential problem with the EFT(OPE) expansion in the singlet,
which is the fact that the coefficients
$|c_{2}^{(\nu=3)}({}^1S_0,{\rm OPE})| = 0.424-0.320$
and $|c_{2}^{(\nu=4)}({}^1S_0,{\rm OPE})| = 0.780-0.137$ are significantly
different from zero, despite containing information about the effective range
(which in principle enters at ${\rm NLO}$ and thus we expect it to have been
fixed at that order).
The numerical mismatch is small though:
the ${\rm N^3LO}$ and ${\rm N^4LO}$ ${}^1S_0$ effective ranges are
$r_0 = (2.71-2.70)\,{\rm fm}$ and $(2.73-2.70)\,{\rm fm}$, respectively,
to be compared with $r_0 = 2.67\,{\rm fm}$ for the Nijmegen II
potential~\cite{PavonValderrama:2005ku} to which we are fitting.
But in terms of power counting expectations the difference is significant:
it reaches a $1.5\%$ and $2.3\%$ at $\nu = 3$ and $4$, to be compared
with $(m_{\pi} / M)^{\nu+1}$, i.e. $0.6\%$ and $0.2\%$
for $M = 0.5\,{\rm GeV}$.
Still, this difference falls within the expected uncertainty
in the extraction of the effective range, i.e.
$r_0 = 2.68(4)\,{\rm fm}$ according to Ref.~\cite{NavarroPerez:2014ovp}.
It should also be noted that a recent implementation~\cite{Thim:2024yks}
of the EFT(OPE) expansion yields results that are closer to
the aforementioned $r_0$ value~\cite{Thim:2024jdv}.

\begin{figure}[ttt]
\begin{center}
  \includegraphics[width=7.50cm]{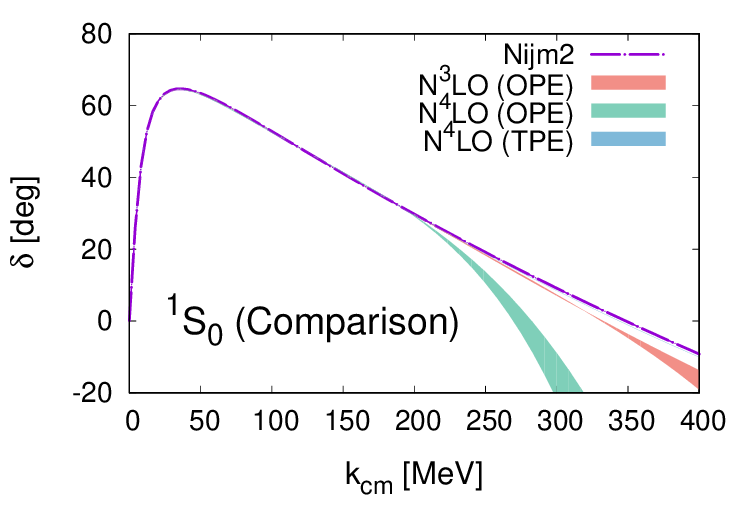}
\end{center}
\caption{Comparison of the phase shifts of the $^1S_0$ partial wave
  in the EFT(OPE) and EFT(TPE) expansions at ${\rm N^3LO}$ and
  ${\rm N^4LO}$, where the bands correspond to varying
  the cutoff in the $R_c = (0.1-0.2)\,{\rm fm}$ range.
  For the EFT(OPE) phase shifts we notice that while at ${\rm N^3LO}$ they
  work really well, this is no longer the case at ${\rm N^4LO}$
  when subleading TPE enters.
  Meanwhile, the EFT(TPE) ${\rm N^4LO}$ phase shift basically overlaps
  with the Nijmegen II ones, and hence is barely visible.
  }
\label{fig:comparison-rctiny}
\end{figure}

In general, with the information contained in Table \ref{tab:coefficients},
it is not possible to reach a clear conclusion of which expansion
is more convergent.
There is however an interesting pattern: for EFT(OPE) the $c^{(\nu)}_{2n}$
coefficients tend to grow in size as the cutoff radius decreases,
particularly in the singlet.
It is tempting to connect this growth with the fact that the EFT(OPE)
expansion fails to reproduce the singlet phase shifts
for $k_{\rm cm} > 200\,{\rm MeV}$ at ${\rm N^4LO}$
with hard cutoffs, as we show in Fig.~\ref{fig:comparison-rctiny}
for $R_c = (0.1-0.2)\,{\rm fm}$.
Yet, this failure is not clearly visible in the singlet $c^{(\nu)}_{2n}$
coefficients, which we show in Table \ref{tab:coefficients-rctiny}.
We can appreciate that $|c_2^{(\nu = 4)}({\rm OPE})| \sim (2.1-2.3)$ and
$|c_4^{(\nu = 4)}({\rm OPE})| \sim (2.6-2.7)$ are somewhat large
for $R_c = (0.1-0.2)\,{\rm fm}$, though
they can be easily rendered of $\mathcal{O}(1)$ by simply
setting $M' = (410-430)\,{\rm MeV}$ (instead of
the original $M' = 500\,{\rm MeV}$),
which would be in line with our previous result
in Eq.~(\ref{eq:M-EFT-comparison}).
This does indeed represent a somewhat reduced expansion radius
with respect to the OPE(TPE) expansion, but it is still very
far away from the $(200-300)\,{\rm MeV}$ momentum range
at which the ${\rm N^4LO}$ phase shifts fail
in Fig.~\ref{fig:comparison-rctiny}.
It is also interesting to notice that $|c_2^{(4)}| >|c_2^{(3)}| > |c_2^{(2)}|$
for the singlet in EFT(OPE), which is concerning as this indicates
corrections to the effective range that are growing
(instead of shrinking) with each order.

However, there is the limitation that the inclusion of subleading TPE is
in a sense a ``unique'' event that happens the moment we reach ${\rm N^4LO}$.
The truth is that the EFT(OPE) expansion is fairly convergent till
${\rm N^3LO}$, and then suddenly deteriorates at the next order,
a pattern which can be appreciated in Fig.~\ref{fig:comparison-rctiny}.
One might wonder how this actually impacts estimations of the breakdown scale
that are based on an order-by-order comparison: if subleading TPE, i.e. the
$Q^3$ piece of the effective finite-range potential, is the reason
for the reduced convergence radius, we will only see its effects at
${\rm N^4LO}$, ${\rm N^8LO}$, ${\rm N^{12}LO}$ and so on (i.e. each
time at which it is iterated again).
Meanwhile, the $c_{2n}^{(\nu)}$ coefficients in between the orders at which
a new iteration of subleading TPE appears, e.g. $\nu=5,6,7$, will most
probably behave as expected.
Indeed, it is perfectly possible that except for a subset of the coefficients
(the ones at $\nu=4,8,12,\dots$) we will end up
with numbers of $\mathcal{O}(1)$.
This is what actually happens to the coefficients
in Table \ref{tab:coefficients-rctiny},
where $|c_{2n}^{(\nu=4)}| \sim (5-10) \, |c_{2n}^{(\nu=3)}|$ for EFT(OPE).
Thus, considerations of naturalness if applied in a naive manner might not
be able to detect the suspected failure of perturbation
theory for subleading TPE.

Finally, it is interesting to notice how the $c_{2}^{(\nu)}$ coefficients of
EFT(TPE) have basically converged in the $R_c = (0.1-0.2)\,{\rm fm}$
cutoff range, indicating a really smooth cutoff dependence.
Indeed, from a comparison of Tables \ref{tab:coefficients} and
\ref{tab:coefficients-rctiny}, it is apparent that the $c_{2}^{(\nu)}$
already converged at $R_c = 0.5\,{\rm fm}$ for the EFT(TPE) singlet.

\begin{table}[thh]
\begin{center}
  \begin{tabular}{|c|c|c|}
    \hline
    \noalign{\smallskip}
    \hline
    {\multirow{2}{*}{Coefficient}} & \multicolumn{2}{c|}{${}^1S_0$} \\
  \cline{2-3}
  & EFT(OPE) & EFT(TPE) \\
  \hline
  $c^{(1)}_0$ & $\mathcal{O}(10^{-3})$ & - \\
  $c^{(2)}_0$ & $\mathcal{O}(10^{-3})$ & - \\
  $c^{(3)}_0$ & $-(1.02-0.86) \cdot 10^{-2}$ & - \\
  $c^{(4)}_0$ & $-(0.107-0.092)$ & $\mathcal{O}(10^{-4})$ \\
  \hline
  $c^{(1)}_2$ & $+(0.878-0.842)$ & - \\
  $c^{(2)}_2$ & $+(0.113-0.053)$ & - \\
  $c^{(3)}_2$ & $+(0.266-0.263)$ & - \\
  $c^{(4)}_2$ & $+(2.33-2.05)$ & $-(0.949-0.950) $ \\
  \hline
  $c^{(3)}_4$ & $-(0.437-0.598)$ & - \\
  $c^{(4)}_4$ & $-(2.70-2.60)$ & -  \\
  \hline \hline
\end{tabular}
\end{center}
\caption{
  Same as Table \ref{tab:coefficients} but for the singlet when calculated
  with the cutoff interval $R_c = (0.1-0.2)\,{\rm fm}$
}
\label{tab:coefficients-rctiny}
\end{table}

\subsection{The non-polynomal contributions to the EFT expansion and their expected size}
\label{app:nonpol}

Here we analyze the scaling properties of the non-polynomial contributions to
the dimensionless function $f^{(\nu)}$, which is basically the rescaled
$\nu$-th order contribution to $\cot{\delta}$,
see Eqs.~(\ref{eq:f-pol-nonpol}) and (\ref{eq:f-def})
for the relevant definitions.
We begin by considering the perturbative integral
\begin{eqnarray}
  \int_0^{\infty} dr \, u_k^{(0)}(r)\,{\left[ V_F(r) u_k(r) \right]}^{(\nu-1)} \, ,
\end{eqnarray}
which, after removing the polynomials and rescaling,
generates ${f}^{(\nu)}_{\rm nonpol}$.
All the terms in ${\left[ V_F u_k \right]}^{(\nu-1)}$ are expected to behave
similarly and thus we might simply focus on the easiest one
to analyze, that is:
\begin{eqnarray}
  \int_0^{\infty} dr \, u_k^{(0)}(r)\,V^{(\nu-1)}_F(r) u_k^{(0)}(r) \, .
\end{eqnarray}
If we take into account that the finite-range potential can be written as
\begin{eqnarray}
  V^{(\nu)}(r) = \frac{2\pi}{\mu}\,\frac{1}{M^{\nu+1}}\,
  \frac{e^{-2m r}}{r^{\nu+3}}\,\left( 1 + \mathcal{O}(Q_{\slashed{k}} r) \right) \, ,
  \nonumber \\
\end{eqnarray}
which is equivalent to Eq.~(\ref{eq:VF-naive}) after the substitution of
$\Lambda_{\rm NN}^{(\nu)}$ by $M$ (plus the assumption that its exponential
decay at long distances is driven by TPE), we arrive to
\begin{eqnarray}
  &&
  \int_0^{\infty} dr \, {\left[u_k^{(0)}(r)\right]}^2\,V^{(\nu-1)}_F(r)
  \nonumber \\
  && \, \sim
  2\pi\,\frac{k}{\mu}\,{\left( \frac{k}{M} \right)}^{\nu}\,
  \int_0^{\infty} \frac{dx}{x^{\nu+2}}\,e^{-\frac{2m}{k} x}\,
  \left( 1 + \mathcal{O}(\frac{Q_{\slashed{k}}}{k} x) \right)\,
  {[u_k^{(0)}]}^2 \, , 
  \nonumber \\ \label{eq:cotd-nu-momentum}
\end{eqnarray}
where we have changed the integration variable to $x = k r$ with the intention
of extracting the power-law dependence with respect to
the center-of-mass momentum $k$.

If we now ignore the expansion in terms of $\frac{Q_{\slashed{k}}}{k} x$
of the finite-range potential, and concentrate in its most
singular term, the non-polynomial part of the perturbative
integral might be extracted after subtracting the first
few terms in the $k^2$ expansion of the square function
\begin{eqnarray}
  \int_0^{\infty} \frac{dx}{x^{\nu+2}}\,e^{-\frac{2 m}{k} x}\,
  \left[ {u_k^{(0)}}^2 - {u_0^{(0)}}^2 - 2 u_0^{(0)} u_2^{(0)} k^2 - \dots \right]
  \, , \nonumber \\
\end{eqnarray}
where for simplicity we have not applied the change of integration
variable (from $r$ to $x=kr$) to the wave function and
its expansion.

We notice here that, after making the subtractions, the perturbative integral
is convergent even if we do not further regularize the EFT potential.

Yet, what matters is the general behavior of the integral
in terms of $2 m / k$ (and the order $\nu$).
For analyzing it, we will momentarily make two simplifying assumptions:
(i) that we are dealing with a pure S-wave and (ii) that the ${\rm LO}$
wave function can be approximated by its asymptotic form, i.e.
\begin{eqnarray}
  u_k^{(0)} \to \cot{\delta}^{(0)} \sin(x) - \cos(x) \, ,
\end{eqnarray}
and (iii) that the part of the perturbative integral that dominates at momenta
$k > 2 m$ is the one corresponding with the $\cos{x}$ part of
the wave function.
This third condition is equivalent to making the substitution
\begin{eqnarray}
  [u_k^{(0)}]^2 \to \cos^2{x} \, ,
\end{eqnarray}
in the perturbative integrals, where subtractions can be trivially be taken
into account by removing terms in the Taylor expansion of $\cos^2{x}$:
\begin{eqnarray}
  &&
  \left( {u_k^{(0)}}^2 - {u_0^{(0)}}^2 - 2\,u_0^{(0)}\,u_2^{(0)}\,k^2 - \dots \right)
  \nonumber \\ 
  && \qquad \qquad \qquad \to \left( \cos^2{x} - 1 + x^2 - \dots  \right) \, .
\end{eqnarray}

Under these conditions and defining the variable $\lambda = 2 m / k$,
we might calculate the first term in the Taylor expansion of
the perturbative integrals in terms of $\lambda$.
If $\nu$ is even, with $2 n = \nu+2$, we find for $n=1,2,3$ ($\nu = 0,2,4$):
\begin{eqnarray}
  && \int \frac{dx}{x^2}\,e^{-\lambda x}\,\left( \cos^2{x} - 1 \right) 
  \nonumber \\ 
  && \qquad = - \frac{\pi}{2} \nonumber + \mathcal{O}\left({\lambda}, {\lambda}\,{\log{\lambda}}\right) \, , \\
  && \int \frac{dx}{x^4}\,e^{-\lambda x}\,\left( \cos^2{x} - 1 + x^2 \right)
  \nonumber \\
  && \qquad = + \frac{\pi}{3} + \mathcal{O}\left( {\lambda}, {\lambda}\,{\log{\lambda}}\right) \, , \\
  && \int \frac{dx}{x^6}\,e^{-\lambda x}\,\left( \cos^2{x} - 1 + x^2 - \frac{x^4}{3}\right)
  \nonumber \\
  && \qquad = - \frac{\pi}{15} \nonumber +
  \mathcal{O}\left( {\lambda}, {\lambda}\,{\log{\lambda}}\right)
  \, ,
\end{eqnarray}
where the general rule for arbitrary even powers is
\begin{eqnarray}
  && \int \frac{dx}{x^{2n}}\,e^{-\lambda x}\,\left( \cos^2{x} - 1 - \sum_{k=1}^{n-1} \frac{(-1)^k 4^k}{2 \,(2k)!}x^{2k}\right) \nonumber \\
  && \qquad = \frac{(-1)^n 4^{n-1}}{(2n-1)!} \frac{\pi}{2} +
  \mathcal{O}\left( {\lambda}, {\lambda}\,{\log{\lambda}}\right) \, .
\end{eqnarray}
It is interesting to notice that when we plug this result
into Eqs.~(\ref{eq:cotd-scaling}) and (\ref{eq:cotd-nu-momentum}),
we generate odd powers of $k / M$, which cannot
originate from a contact-range potential.

If $\nu$ is odd, with $2n+1 = \nu+2$, for $n=1,2,3$ ($\nu=1,3,5$) we find
\begin{eqnarray}
  && \int \frac{dx}{x^3}\,e^{-\lambda x}\,\left( \cos^2{x} - 1 + x^2  \right)
  \nonumber \\
  && \qquad = - \log{\lambda} +
  \mathcal{O}\left( {\lambda}, {\lambda}\,{\log{\lambda}}\right)
  \, , \\
  && \int \frac{dx}{x^5}\,e^{-\lambda x}\,\left( \cos^2{x} - 1 + x^2 - \frac{x^4}{3} \right)
  \nonumber \\
  && \qquad = +\frac{1}{3}\,\log{\lambda} +
  \mathcal{O}\left( {\lambda}, {\lambda}\,{\log{\lambda}}\right)
  \, , \\
  && \int \frac{dx}{x^7}\,e^{-\lambda x}\,\left( \cos^2{x} - 1 + x^2 - \frac{x^4}{3} + \frac{2 x^6}{45} \right) 
  \nonumber \\
  && \qquad = -\frac{2}{45}\,\log{\lambda} +
    \mathcal{O}\left( {\lambda}, {\lambda}\,{\log{\lambda}}\right)
  \, ,
\end{eqnarray}
where the general rule for arbitrary $n$ is
\begin{eqnarray}
  && \int \frac{dx}{x^{2n+1}}\,e^{-\lambda x}\,\left( \cos^2{x} - 1 - \sum_{k=1}^{n} \frac{(-1)^k 4^k}{2 \,(2k)!} x^{2k}\right) \nonumber \\
  && \qquad = \frac{(-1)^n 4^n}{2 \,(2n)!}
  \log{\lambda} + \mathcal{O}\left( {\lambda}, {\lambda}\,{\log{\lambda}}\right) \, .
\end{eqnarray}
By plugging the results above into Eqs.~(\ref{eq:cotd-scaling}) and
(\ref{eq:cotd-nu-momentum}),
we generate even powers of $k / M$, but multiplied by $\log{(k / 2 m)}$,
which again indicate that we are not dealing with a contact-range potential.

Even though the previous analysis is rudimentary in nature,
it paints a picture in which the non-polynomial part of
the perturbative integral scales as $\mathcal{O}(1/\nu!)$
as the order $\nu$ of the EFT expansion is increased.
If this were to be the case, and putting the pieces that we have together,
for $\nu$ even we find
\begin{eqnarray}
  f^{(\nu)}_{\rm nonpol}(\frac{k}{Q_{\slashed{k}}}) &\sim& 2\pi \, \frac{M}{\mu} \,
  \left( \frac{k}{Q_{\slashed{k}}} \right)^{\nu+1}\,
  \frac{(-1)^{\nu/2+1} 2^{\nu}}{(\nu+1)!} \frac{\pi}{2}  \nonumber \\
  &\times& \left[ 1 + \mathcal{O}\left( \frac{Q_{\slashed{k}}}{k} , \frac{Q_{\slashed{k}}}{k}\,\log{\left( \frac{Q_{\slashed{k}}}{k} \right)} \right)\,\right]  \nonumber \\
  &\sim& \pi^2\,\frac{2^{\nu}}{(\nu+1)!} \, ,
  \label{eq:f-nu-even}
\end{eqnarray}
while for $\nu$ odd we arrive at
\begin{eqnarray}
  f^{(\nu)}_{\rm nonpol}(\frac{k}{Q_{\slashed{k}}}) &\sim& 2\pi \, \frac{M}{\mu} \,
  \left( \frac{k}{Q_{\slashed{k}}} \right)^{\nu+1}\,\log{(\frac{k}{Q_{\slashed{k}}})}\,
  \frac{(-1)^{(\nu+1)/2} 2^{\nu}}{2\sqrt{2}\,(\nu+1)!}  \nonumber \\
    &\times& \left[ 1 + \mathcal{O}\left( \frac{Q_{\slashed{k}}}{k} , \frac{Q_{\slashed{k}}}{k}\,\log{\left( \frac{Q_{\slashed{k}}}{k} \right)} \right)\,\right]  \nonumber \\
  &\sim& \frac{\pi}{\sqrt{2}}\,\frac{2^{\nu}}{(\nu+1)!} \, .
  \label{eq:f-nu-odd}
\end{eqnarray}
Of course a more complete analysis should evaluate the size of
the additional terms in the potential as well as terms
stemming from higher order perturbation theory.
Yet, provided that the number of these terms do not scale as $\nu!$,
the conclusion is that
$f^{(\nu)}_{\rm nonpol} \sim \mathcal{O}(2^\nu \pi / \nu!)$.
That is, the non-polynomial pieces tend to generate contributions that
are convergent within the EFT expansion and thus they do not provide
information about the radius of convergence of the EFT.

\subsection{Comparison with the usual Bayesian analysis}
\label{app:bayesian}

Here we briefly discuss how to incorporate the previous analysis about
the scaling properties of perturbative amplitudes
into Bayesian techniques, as well as the potential differences
with the more standard approaches currently used
in the literature~\cite{Schindler:2008fh,Wesolowski:2015fqa,Furnstahl:2015rha,Melendez:2017phj,Wesolowski:2018lzj,Melendez:2019izc}.

Usually, the starting point of most Bayesian analyses is a series in
the form~\cite{Furnstahl:2015rha}
\begin{eqnarray}
  y_{k}(x) = y_{\rm ref}\,\sum_{n =0}^{k} c_n(x)\,Q^{n}(x) \, ,
  \label{eq:generic-exp}
\end{eqnarray}
where $y_k(x)$ is the observable quantity we are interested in, $y_{\rm ref}$
a reference value for this quantity (which serves the purpose of removing
the dimensions), $c_n(x)$ a dimensionless coefficient, $Q(x)$
the expansion parameter and $k$ the highest order
we are considering in the expansion.
The problem we might want to solve could be for instance
how to make a reliable estimation of the breakdown scale (which is related to
the expansion parameter $Q(x)$) given the assumption that
the $c_n(x)$ coefficients are natural,
though the application of Bayesian techniques is obviously
not limited to this example.
Here what we will discuss instead is how to improve over
the previous ansatz for the EFT expansion.

If we now apply Eq.~(\ref{eq:generic-exp}) to the phase shift:
\begin{eqnarray}
  \delta^{{\rm N^{\nu_{\rm max}}LO}}(k) = \delta_{\rm ref}\,\sum_{\nu = 0}^{\nu_{\rm max}}\,\hat{\delta}^{(\nu)}(k)\,{\left( \frac{Q}{M}\right)}^{\nu} \, ,
  \label{eq:naive-bayesian}
\end{eqnarray}
where $\delta_{\rm ref}$ is a reference value of the phase shift and
$\hat{\delta}^{(\nu)}$ a rescaled phase shift in which
we have removed the $(Q/M)$ factors.
$Q$ here is usually interpreted as a function of the light scales in the system,
$k$ and $m_{\pi}$, e.g. $(k^n + m_{\pi}^n)/(k^{(n-1)} + m_{\pi}^{(n-1)})$ with $n$
an integer number~\cite{Melendez:2017phj}.

Writing the EFT expansion in the previous way is completely correct.
Yet, we might instead try to exploit the scaling properties
we have derived from analyzing the subleading contributions.
If we use again the phase shift as the example, in a first approximation
we would obtain
\begin{eqnarray}
  \delta^{{\rm N^{\nu_{\rm max}}LO}}(k) = \delta^{(0)} + \frac{2\mu}{k}\,\sum_{\nu=1}^{\nu_{\rm max}}\,g^{(\nu)}(\frac{k}{Q_{\slashed{k}}})\,{\left( \frac{Q_{\slashed{k}}}{M}\right)}^{\nu+1} \, , \nonumber \\
\end{eqnarray}
which can be further refined by taking into account the existence of polynomial
and non-polynomial terms
\begin{eqnarray}
  && \delta^{{\rm N^{\nu_{\rm max}}LO}}(k) \nonumber \\
  && \quad = \delta^{(0)} + \frac{2\mu}{k}\,\sum_{\nu=1}^{\nu_{\rm max}}\,\left[ g^{(\nu)}_{\rm pol} + g^{(\nu)}_{\rm nonpol} \right]\,{\left( \frac{Q_{\slashed{k}}}{M}\right)}^{\nu+1} \, , \nonumber \\
  && \quad = \delta^{(0)} - \frac{2\mu}{k}\,\sin^2{\delta^{(0)}}\,
  \sum_{\nu=1}^{\nu_{\rm max}}\,\left[ \sum_{n=0}^{n_{\rm max}(\nu)} c_{2n}^{(\nu)}{\left( \frac{k}{Q_{\slashed{k}}}\right)}^{2n} \right]\,{\left( \frac{Q_{\slashed{k}}}{M}\right)}^{\nu+1} \nonumber \\
  && \quad + \sum_{\nu=1}^{\nu_{\rm max}}\,g^{(\nu)}_{\rm nonpol}\,{\left( \frac{Q_{\slashed{k}}}{M}\right)}^{\nu+1} \, , \label{eq:EFT-bayesian}
\end{eqnarray}
where the sum of $g^{(\nu)}_{\rm nonpol}$ contributions is suspected to be always
convergent (or at least their size is not of $\mathcal{O}(1)$),
and hence they might be ignored.
With this new expansion, the problem is now reduced to determining
the probability of the breakdown scale $M$ given the assumption
that the $c_{2n}^{(\nu)}$ coefficients obtained from this expansion
are of natural size.

The main advantage of using the expansion derived from the scaling
properties of the EFT amplitudes, i.e. Eq.~(\ref{eq:EFT-bayesian}),
is that we do not need the ad-hoc interpolators for $(Q/M)^{\nu}$
that we would need in the naive expansion
of Eq.~(\ref{eq:naive-bayesian}).
The rationale for interpolators such as $Q \to (k^n + m_{\pi}^n)/(k^{(n-1)} + m_{\pi}^{(n-1)})$ is that we are not sure what is the actual role played by the momentum
and the pion mass (or other light scales we might want to add)
in the EFT expansion.
The analysis of Appendix \ref{app:analysis}, which resulted
in Eq.~(\ref{eq:EFT-bayesian}), effectively solves this problem.
Yet, there is an important disadvantage though: with Eq.~(\ref{eq:EFT-bayesian})
we now have an integer (and small) number of coefficients,
instead of coefficients on a continuous variable (e.g. the momentum).
This somewhat limits the applicability of Bayesian statistics, as there is
a smaller set of ${\mathcal O}(1)$ quantities from which to extract
information about the breakdown scale.


%

\end{document}